\documentclass[12pt]{article}
\usepackage{graphicx}
\usepackage[margin=1.25in]{geometry}
\usepackage[usenames,dvipsnames]{color}
\usepackage{url}
\usepackage{caption,subcaption, cite}
\usepackage[colorlinks = true,
            linkcolor = blue,
            urlcolor  = blue,
            citecolor = blue,
            anchorcolor = blue]{hyperref}

\newcommand\pubnumber{  KEK Preprint 2017--23\\
 SLAC--PUB--17130}
\newcommand\pubdate{August, 2017}

\def\SLAC{SLAC,
    Stanford University, Menlo Park, CA 94025, USA}

\def\kek{High Energy Accelerator Research Organization (KEK), Tsukuba,
  Ibaraki, JAPAN  }
\def\Tokyo{ICEPP, University of Tokyo, Hongo, Bunkyo-ku, Tokyo,
  113-0033, JAPAN}
\def\SNU{Dept. of Physics and Astronomy, Seoul National
  Univ.,
Seoul 08826, KOREA}

\def\Title#1{\begin{center} {\Large #1 } \end{center}}
\def\Author#1{\begin{center}{ \sc #1} \end{center}}

\def\submit#1{\begin{center}Submitted to {\sl #1} \end{center}}
\newcommand\pubblock{\rightline{\begin{tabular}{l} \pubnumber\\
         \pubdate \end{tabular}}}
\newenvironment{Abstract}{\begin{quotation} \begin{center}
                       ABSTRACT
     \end{center}\bigskip  }{\end{quotation}}

\def\submit#1{\begin{center}Submitted to {\sl #1} \end{center}}
\def\Acknowledgements{\bigskip  \bigskip \begin{center} \begin{large}
             \bf ACKNOWLEDGEMENTS \end{large}\end{center}}

\def\beq{\begin{equation}}
\def\eeq#1{\label{#1}\end{equation}}
\def\eeqn{\end{equation}}

\newenvironment{Eqnarray}%
   {\arraycolsep 0.14em\begin{eqnarray}}{\end{eqnarray}}
\def\beqa{\begin{Eqnarray}}
\def\eeqa#1{\label{#1}\end{Eqnarray}}
\def\eeqan{\end{Eqnarray}}
\def\CR{\nonumber \\ }

\def\leqn#1{(\ref{#1})}

\let\bar=\overbar

\def\etal{{\it et al.}}

\def\VEV#1{\left\langle{ #1} \right\rangle}

\def\lsim{\mathrel{\raise.3ex\hbox{$<$\kern-.75em\lower1ex\hbox{$\sim$}}}}
\def\gsim{\mathrel{\raise.3ex\hbox{$>$\kern-.75em\lower1ex\hbox{$\sim$}}}}

\def\L{{\cal L}}
\def\M{{\cal M}}

\def\L{{\cal L}}

\def\half{\frac{1}{2}}
\def\thalf{\frac{3}{2}}

\def\del{\partial}
\def\Dslash{\not{\hbox{\kern-4pt $D$}}}
\def\dslash{\not{\hbox{\kern-2pt $\del$}}}

\def\Dlr{\mathrel{\raise1.5ex\hbox{$\leftrightarrow$\kern-1em\lower1.5ex\hbox{$D$}}}}

\def\ee{e^+e^-}

\def\msb{{\bar{\scriptsize M \kern -1pt S}}}

\def\drb{{\bar{\scriptsize D \kern -1pt R}}}

\def\eps{\epsilon}

\makeatletter
\def\section{\@startsection{section}{0}{\z@}{5.5ex plus .5ex minus
 1.5ex}{2.3ex plus .2ex}{\large\bf}}
\def\subsection{\@startsection{subsection}{1}{\z@}{3.5ex plus .5ex minus
 1.5ex}{1.3ex plus .2ex}{\normalsize\bf}}
\def\subsubsection{\@startsection{subsubsection}{2}{\z@}{-3.5ex plus
-1ex minus  -.2ex}{2.3ex plus .2ex}{\normalsize\sl}}

\renewcommand{\@makecaption}[2]{%
   \vskip 10pt
   \setbox\@tempboxa\hbox{\small #1: #2}
   \ifdim \wd\@tempboxa >\hsize     
       \small #1: #2\par          
     \else                        
       \hbox to\hsize{\hfil\box\@tempboxa\hfil}
   \fi}

\makeatother

\begin{document}
\begin{titlepage}
\pubblock

\vfill
\Title{Model-Independent Determination of the Triple Higgs Coupling at
$\ee$ Colliders}
\vfill
\Author{Tim Barklow$^a$, Keisuke Fujii$^b$, Sunghoon
  Jung$^{ac}$,   \\ Michael
  E. Peskin$^{a}$, and  Junping Tian$^d$}
\bigskip
\begin{center} { \it
$^a$  \SLAC \\ 
$^b$  \kek \\
$^c$  \SNU \\ 
$^d$   \Tokyo }
\end{center}
\vfill
\begin{Abstract}
The observation of Higgs pair production at high-energy colliders
can give evidence for the presence of a triple Higgs coupling.
However, the actual determination of the value of this coupling
is more difficult.  In the context of general models for new
physics, double Higgs production processes can receive contributions from
many possible beyond-Standard-Model effects.  This dependence must be
understood if one is to make a definite statement about the deviation of
the Higgs field potential  from the Standard Model.   In this paper, we
study the extraction of the triple Higgs coupling from the process
$\ee \to Z h h$.   We show that, by combining the measurement of this
process with other measurements available at a 500~GeV $\ee$ collider,
it is possible to quote model-independent limits on the Effective Field Theory parameter
$c_6$ that parametrizes modifications of  the Higgs potential.   We present precise error estimates
based on the anticipated ILC physics program, studied with full
simulation.   Our analysis also gives new insight into the
model-independent extraction of the Higgs boson coupling  constants
and total width  from $\ee$  data.
\end{Abstract}
\vfill
\submit{Physical Review D}
\vfill

\newpage
\tableofcontents
\end{titlepage}

\def\thefootnote{\fnsymbol{footnote}}
\setcounter{footnote}{0}

\section{Introduction}

The discovery of the Higgs boson in 2012~\cite{ATLAS,CMS} closed one
chapter in our understanding of elementary particle physics but opened
another.   The observation of this particle, at a relatively low value
of the mass and with large couplings to $W$, $Z$, and heavy fermions,
confirmed the picture of the Standard Model that masses originate in
the vacuum expectation value of a scalar field.  At the same time,
this observation deepened the mysteries associated with this particle,
and also offered a path to solving these mysteries through precision
measurements of the Higgs boson properties.

The goal of this paper is to  develop methods for the precision extraction of
Higgs boson couplings using Effective Field Theory to represent the
most general effects of new physics on the Higgs boson.   Effective
Field Theory (EFT) has been applied to the theory of a Higgs boson in many
papers, for example, \cite{SILH,newSILH,Maxim} and has been adopted as
a canonical framework for analyzing Higgs boson measurements at the
LHC~\cite{Handbookfour}. Still, we feel that the full power of this
formalism is not appreciated.  The reason for this is that a
fully general treatment of EFT brings in a very large number of free
parameters.  It has not been clear how to constrain all of these
parameters simultaneously from experimental measurements~\cite{Pomarol}.

In this paper, we will show that this problem can be solved by making
use of the large number of observables that can be measured with high
precision at future $\ee$ colliders.  In our analysis, we analyze
the extension of the Standard Model (SM) by addition of 10 effective
operators that describe the most general new physics effects on 
the couplings of the Higgs boson to the $W$, $Z$, and $\gamma$ and 
the light leptons.   We  show how to determine the coefficients of
these operators
systematically. We apply this method to solve an important problem
involving the measurement of the Higgs boson self-coupling.   We 
also present formulae that extend this method to a general,
model-independent approach to the extraction of Higgs boson couplings
from data.

The specific aim of this paper is to solve the following problem for
the Higgs boson self-coupling:  The Higgs boson self-coupling is 
predicted by the SM.  The experimental test of this
prediction is of great importance both for our basic understanding of
electroweak symmetry breaking and for the linkage of this issue to other
questions such as Higgs CP violation and electroweak baryogenesis~\cite{EWbaryogenesis}.
 If the SM were exact except for a perturbation that
changes the triple-Higgs coupling, it would be possible to measure
this coupling by observing a 
change in the rate of double Higgs production.  This measurement has
been studied in some detail for $gg\to
hh$ at hadron colliders~\cite{ATLAShh,CMShh,FCChh}, for 
$\ee\to Z hh$~\cite{ILChh}  and for the vector boson fusion processes
 $\ee \to \nu\bar \nu hh$~\cite{ILChh,CLIChh}, and $ud \to du
 hh$\cite{FCChh}.  In some cases, future experiments would be highly
 sensitive to a deviation of the rate from the SM
 prediction.     A series of papers, beginning with \cite{McCullough},
 have even discussed extracting the triple-Higgs coupling from single
 Higgs production measurements, through its effects in  loop diagrams.
However, the assumption that triple-Higgs coupling is altered by some
effect of new physics while all other Higgs boson couplings remained
unchanged is extremely artificial.   It is more likely that new physics
alters many  of the couplings of the Higgs boson and alters the rate
of single and double Higgs production through many different vertices.   But how,
then, can we distinguish the effects of changes in the  Higgs boson potential from
perturbations induced by other new physics effects?

This question has hardly been studied in the literature, and its
resolution is not  straightforward.   The paper \cite{Continoandcrew}
studied the influence of a second operator perturbation of the
SM  and shows that this effect can be distinguished from a
change in the triple Higgs coupling
by studying the dependence of the
double Higgs production cross section on $m(hh)$ at center of mass
energies well above threshold.   The 
papers~\cite{Goertz,Azatov,Carvalho} studied the process $gg\to hh$ at proton
colliders and suggested measurements beyond the total cross section 
measurements that discriminate contributions of different operators.
The paper~\cite{Global} studied the discrimination of loop effects of
the triple Higgs coupling in single-Higgs processes from other EFT effects.
   In all of these cases, the extension of the method to
high precision and to general new
physics perturbations seems very challenging.  

The best way to attack this problem is to enumerate all possible new
physics effects that influence the cross section for double Higgs
production and to constrain them one by one in a systematic way,
leaving, at the end, only the triple Higgs coupling as a free
parameter.  In this paper, we explain how to do that through the
use of the EFT parametrization of possible deviations from the
SM.   We concentrate on the extraction of the triple
Higgs coupling from the rate of the reaction  $\ee\to
Z hh$, which can be measured already at a 500~GeV $\ee$ collider.
Effects on the $\ee\to Zhh$ cross section from dimension-6 EFT
operators have been studied previously in~\cite{Shaouly}.

  Our
analysis will involve a total of 17 EFT operator coefficients.  Of
these,  one is the parameter $c_6$ that shifts the triple Higgs
coupling,
9 others  govern the couplings among vector boson, leptons, and the
Higgs boson, while
another 7 appear in other Higgs decay amplitudes that will enter our
analysis.  This seems at first sight extremely complex, but we will
see that each coefificient has its place and can be constrained in a
physically apparent way.

The outline of this paper is as follows:  In Section 2, we set up our
formalism
for the EFT analysis of Higgs and vector boson process.  We  present 
a basic strategy for our analysis by writing the
the  potentially measureable vector boson, lepton, and Higgs couplings 
in terms of EFT coefficients.   We justify the restriction to this
parameter set and  discuss some approximations we make to simplify 
the analysis.  And, we present our method for including
 the constraints on the
EFT coefficients coming from precision electroweak measurements.
In Section 3, we present the constraints from measurements 
of $\ee\to W^+W^-$ at future $\ee$ colliders and describe these 
constraints quantitatively using the results of full-simulation studies
for the International Linear Collider (ILC).  This process has prevoiusly
been analyzed in an EFT formalism, 
using LEP and LHC results, in \cite{Riva,Rivatwo}.   In Section 4, we 
discuss the effect of the expected measurements of Higgs branching
ratios to $\gamma\gamma$, $\gamma Z$, and $\mu^+\mu^-$  at the LHC. 

 In Section 5, we
explain how the measurement of the cross section, angular
distribution, and polarization asymmetry for $\ee\to Zh$ constrain the
EFT parameters.  An EFT analysis of the total cross section for
this process has previously been given in \cite{Maxim}.   We will show
that these measurements supply the missing pieces of information
needed to constraint the full set of 9 operators responsible for new
physics effects in vector boson, lepton, and Higgs couplings.

In principle, this should be enough information to extract the triple
Higgs coupling parameter $c_6$ from the measurement of the cross
section for $\ee\to Zhh$.   However, in practice, the constraint turns
out not to be strong enough.   We can find additional
constraints on the EFT parameters by studying the other major
single-Higgs production process available at $\ee$ colliders, the 
$W$ fusion process $\ee\to \nu\bar\nu h$.  This
 reaction has a larger cross section than $\ee\to Zh$ at
500~GeV, and it also depends strongly on the EFT parameters.
However, in this case, there is no specific Higgs boson tag and so the cross
section cannot be measured in model-independent way.  To make use of
this process, we will need also to study the Higgs decay partial widths.
These also have expansions in EFT parameters.  These bring in another
7 parameter beyond our original set, but in the end, all of the
parameters can be strongly constrained.

Thus, in 
Section 6, we work out formulae in terms of EFT parameters 
 for the total cross section for $\ee\to \nu\bar \nu h$.  In 
  Section 7, we present the EFT formulae for the various Higgs
  boson partial widths.    This formalism provides the basis not only
  to determine the shift of the triple Higgs coupling but also to
  develop a method for determining the full set of Higgs boson
  couplings in model-independent way.  The
  implications of this formalism for Higgs coupling determination at
  $\ee$ colliders
 will be presented in a companion paper~\cite{ImpHiggs}. 

Finally, in Section 8, we present
the dependence of the cross section for $\ee\to Zhh$ in terms
of our full set of parameters.  We estimate the error on the
  prediction of the 
total cross section for this reaction due to uncertainties from all
new physics effects except for the variation of the triple Higgs
coupling.  This estimate makes use of projections for the accuracy of
high-precision measurements of single-Higgs processes expected to be
carried out at the ILC~\cite{ILCHiggsWhite,ILCCase,ILCscenarios}.    We
estimate  that this  uncertainty in the total cross section will be
2.4\%, corresponding to a 5\% systematic uncertainty
 in the determination of the triple Higgs
coupling.  This is attractively small and should be subdominate to expected statistical
and direct experimental systematic errors. 
 
Section 9  presents our conclusions.   Appendix A summarizes the
formulae used in our fit.  Appendix B specifies the inputs to the fit
in more detail.

\section{Effective Field Theory formalism}

In this paper, we  represent the effects of new physics by writing
an extension of the SM  as an Effective Field Theory.   The
SM is already the most general theory with operators
of dimension 4 or lower,
$SU(3)\times SU(2)\times U(1)$ gauge invariance,  the known spectrum
of quarks and leptons, and one $SU(2)$-doublet Higgs field.   If new
physics effects are due to new heavy particles of mass at least $M$,
their effects can be represented by adding operators of dimension 6.
The effects of these operators are suppressed by factors $1/M^2$.
  For $M > 500$~GeV, as suggested
by LHC results, these factors already push the size of the most
general new physics effects below the current sensitivity of LHC Higgs
measurements.   Effects of operators of dimension 8 and higher are
suppressed by additional powers of $1/M^2$, and we will neglect
them in this discussion.

The restriction to dimension 6 operator perturbations leaves a great
deal of freedom.
For the SM with one fermion generation, there are a total
of 84 independent dimension 6 operators that can be added to the
Lagrangian.  Of these, 8 are baryon-number violating and, of the
remainder,  59 are CP-conserving while 17 are
CP-violating~\cite{Polish,Melia}.  Fortunately, not all of these
operators contribute to the processes of interest in a given study.
For the goals of this paper, a subset of 17 of these operators will
suffice for a general analysis. These divide into a set of 10 govering
vector boson, lepton, and Higgs boson couplings and another set of 7,
which will be introduced in Section 7, needed to other possible Higgs decays.

\subsection{Operator basis}

One aspect of the study of the dimension-6 effective operators is
that there are many possible choices of basis.    In this paper, we
will study processes that involve only light leptons, electroweak gauge
bosons, and Higgs bosons.  Thus, we should choose an operator basis
that is convenient for analyzing this particular system. 
We choose a  basis that
includes the minimum number of operators that include only gauge
fields and Higgs fields, using the equations of motion to convert
purely bosonic operators to operators that include 
include quark and lepton fields.  Some operators that involve the
lepton fields must also be included in the analysis.  The use of equations of motion to make
these reductions and other aspets of EFT formalism are
 explained in \cite{Polish,Wells,Rosetta,Murayama,Trott} and
many other papers.  A very convenient choice for our 
analysis is the ``Warsaw'' basis put forward  in \cite{Polish}.  In
the CP-conserving case, this basis contains only 7 operators
containing only $W$, $Z$, and Higgs boson field, and another 3
relevant operators
containing lepton fields.   We
will slightly rearrange the pure Higgs operators, as is done in the
``SILH'' basis~\cite{SILH,newSILH}, for convenience in the analysis.   In the
CP-violating case, another 4 operators need to be included.

In this section, we will present the basic formalism and notation for
these operators.   We generally follow the conventions of 
\cite{Veronica}, which in turn are based on~\cite{SILH,newSILH}.  The
same basis is used (with slightly different notation) in
\cite{Maxim}. 

Our analysis will use 10 CP-conserving operators from the  Warsaw
basis of dimension-6 operators.
 We will notate these as 
\beqa
\Delta \L &=& {c_H\over 2 v^2} \del^\mu(\Phi^\dagger \Phi)
\del_\mu(\Phi^\dagger \Phi) 
 +  {c_T\over 2v^2} ( \Phi^\dagger \Dlr{}^\mu \Phi)  ( \Phi^\dagger \Dlr_\mu \Phi)
- {c_6\lambda \over v^2} (\Phi^\dagger \Phi)^3 \CR
&  & + {g^{2} c_{WW} \over m_W^2} \Phi^\dagger \Phi
W^a_{\mu\nu} W^{a\mu\nu} +{4 gg^{\prime} c_{WB} \over m_W^2}
\Phi^\dagger t^a \Phi
W_{\mu\nu}^a B^{\mu\nu} \CR & & + {g^{\prime 2} c_{BB} \over m_W^2} \Phi^\dagger \Phi
B_{\mu\nu} B^{\mu\nu} 
 + {g^3 c_{3W}\over m_W^2} \eps_{abc} W^a_{\mu\nu} W^{b
  \nu}{}_\rho W^{c \rho \mu} \CR
& & + i {c_{HL}\over v^2}  ( \Phi^\dagger \Dlr{}^\mu \Phi) (\bar L
\gamma_\mu L)  + 4 i {c'_{HL}\over v^2}  ( \Phi^\dagger t^a \Dlr{}^\mu \Phi) (\bar L
\gamma_\mu t^a L) \CR & & +  i {c_{HE}\over v^2}  ( \Phi^\dagger \Dlr{}^\mu \Phi) (\bar e
\gamma_\mu e) \ .
\eeqa{firstL}
The parameter $c_6$ shifts the Higgs potential.  The other parameters
express different possible new physics effects.  The operators in
  \leqn{firstL} must be defined at a specific momentum scale.  We take
  this scale to be close to 500~GeV.

The 4 dimension-6  CP-violating operators can be written as 
\beqa
\Delta \L_{CP} &=&+ {g^{2} {\tilde c}_{WW} \over m_W^2} \Phi^\dagger \Phi
W^a_{\mu\nu} {\widetilde W}^{a\mu\nu} +{4 gg^{\prime}  {\tilde c}_{WB} \over m_W^2}
\Phi^\dagger t^a \Phi
W_{\mu\nu}^a  {\widetilde B}^{\mu\nu} \CR & & 
+ {g^{\prime 2}  {\tilde c}_{BB} \over m_W^2} \Phi^\dagger \Phi
B_{\mu\nu}  {\widetilde B}^{\mu\nu} 
 + {g^3  {\tilde c}_{3W}\over m_W^2} \eps_{abc} W^a_{\mu\nu} W^{b
  \nu}{}_\rho  {\widetilde W}^{c \rho \mu} 
\eeqa{CPL}

Operators involving gluon fields are not needed for our
analysis, and so do not appear in \leqn{firstL} and \leqn{CPL}. 
All of the parameters $c_i$ and $\tilde c_i$ are dimensionless.

Other bases for the dimension-6  operators include additional bosonic
operators called ${\cal O}_W$ and ${\cal  O}_B$.    When these operators
are eliminated using the equations of motion, the operators 
 ${\cal O}_{HL}$, ${\cal O}'_{HL}$, and ${\cal O}_{HE}$, with a Higgs
 current and a lepton current, are generated.   These  terms containing
 lepton fields play an surprisingly important role in our analysis.
 They cannot be ignored. 

The notation of these equations requires some explanation.
$W^a_{\mu\nu}$ and $B_{\mu\nu} $ are the Yang-Mills field strength
tensors for $SU(2)$ and $U(1)$.  $D_\mu$ is the gauge-covariant
derivative, $t^a = \sigma^a/2$, and 
\beqa
     \Phi^\dagger \Dlr_\mu \Phi &=&  \Phi^\dagger D_\mu \Phi - D_\mu
     \Phi^\dagger \Phi  \CR
    \Phi^\dagger t^a \Dlr_\mu \Phi &=&  \Phi^\dagger  t^a D_\mu \Phi - D_\mu
     \Phi^\dagger t^a \Phi  
\eeqa{Darrowdef}
The tilded field strengths in \leqn{CPL} are
\beq 
   \widetilde W^a_{\mu\nu} = \half \eps_{\mu\nu\lambda\sigma}
   W^{a\lambda\sigma} \ , \quad 
  \widetilde B_{\mu\nu} = \half \eps_{\mu\nu\lambda\sigma}
  B^{\lambda\sigma}  \ .
\eeqn

Finally, we will write   
\beq
   s_w^2 = \sin^2\theta_w = {g^{\prime 2}\over g^2 + g^{\prime 2}}\ , \quad
  c_w^2 = \cos^2\theta_w = {g^2\over g^2 + g^{\prime 2}} \ , 
\eeqn
in terms of the $SU(2)\times U(1)$ couplings in the Lagrangian.
The renormalization prescription that we will use for
the couplings and 
the weak mixing angle will be given in Section 2.3 below.

\subsection{Simplifications}

Our analysis will include a number of simplications that we will now
enumerate.  None of these simplifications has a significant effect on
our final answers.  We will explain how the analysis given here can be
systematically improved to relax some of these simplifying
assumptions.

First, we will work at the tree level and strictly to linear order in
the dimension-6 operator coefficients.  For the central values in the
fit, we will assume that the SM is precisely valid and the EFT
coefficients in
\leqn{firstL} are zero.

We will primarily be concerned with the sensitivity of $\ee$
experiments to values of the $c_i$
 of 1\% and below.  It is possible that some of the $c_i$ could be
 larger, even of order 1, consistent with current data.   That would
 be a wonderful situation. But in this paper  we are trying to probe the limits
 of sensitivity of future experiments. We estimate corrections to the
 linear approximation in the $c_i$  in a manner consistent with this
 viewpoint.   That is, effects quadratic in the $c_i$ should be of order
 $ 10^{-4}$,  effects due to operators of dimension 8 should be of order
$ 10^{-4}$, and effects of electroweak radiative corrections to the
 terms linear in the $c_i$, including operator mixing of dimension-6
 operators, should be of order $\alpha_w/\pi \cdot 1\%$.  Then we may
 neglect effects of all three of these types in the analysis 
presented here.   

It has been pointed out in \cite{Shaouly} that some
terms of order $c_I$ are enhanced by factors  proportional to
$s/m_Z^2$.  We will see such enhancements appearing in our analysis.
However, we will see that 
the uncertainties on the corresponding EFT coefficients are
extremely small, such that the corrections that include
these enhancement factors are still restricted to be of order 1\%.  Explicit examples
will be discussed in Section~5 and Section~8.

Since we are concerned with such small corrections to the Standard
Model predictions for Higgs cross sections and decay rates,
comparisons to data should use high-precision Standard Model
calculations of the cross sections and rates.  In this paper, we will
compute Standard Model rates only to the tree level.  This will
suffice for estimating the sensitivity to the new physics corrections
that we consider in this paper.   An actual  experimental analysis
will need to combine our formulae with Standard Model predictions
computed at least to 1-loop order, and, in most cases, to 2-loop
order, in 
electroweak corrections. 

Our analysis will involve all coefficients in \leqn{firstL} other
  than the coefficient $c_6$ that shifts the triple Higgs coupling.
In this analysis, we  will not need to assume that $c_6$ is small, though we will
  ignore effects of $c_6$ in loop diagrams (proportional to $\lambda^2
  c_6/(4\pi)^2$).   This is important to note, because models of
  electroweak baryogenesis expect values of $c_6$ of order 1~\cite{EWbaryogenesis}, and the            
expected error on $c_6$ from ILC is 27\%~\cite{ILChh,Duerig}. 
 If $c_6$ were indeed of
  order 1\%, along with the other $c_i$ coefficients, its effect would
  not be measurable at the ILC or at any other proposed collider.
 One might ask
if it is consistent to have $c_6$ of order 1 while the other EFT
coefficients are extremely small.   Some examples of models with this
property
are given in \cite{Azatov,Endo,Hashino}, and in
 Section 2.3 of \cite{Global}.  More
likely, corrections from the new physics that modifies $c_6$ will also
shift the parameters such as $c_H$ and $c_{WW}$ in \leqn{firstL} to
nonzero values of order a few percent.  These shifts might be the
first indication of a correction to the Higgs sector Lagrangian.   The
shifts will not affect our error estimate for $c_6$, though they will
of course alter the value of $c_6$ that is extracted from the cross
section for double Higgs production. 

Second, we will ignore some possible dimension-6 operator corrections
involving the light leptons.   We will consider the three coefficients
$c_{HL}$, $c^\prime_{HL}$, $c_{HE}$ as 
independent free parameters. 
Taking this presciption,  we are explicitly not assuming that the dimension-6
corrections are ``oblique'' (in the language of \cite{PT}) or
``universal'' (in the language of \cite{Wells}).   However, we  will assume
electron-muon-tau  universality.  We will need, first, the constraint  $c'_{HL} (\mu) =
c'_{HL}(e)$, to use $G_F$ together with constraints from precision
electroweak measurements involving electrons only.   This assumption can be tested by measuring the equality
of the $W$ boson branching ratios to $\mu$ and $e$ using the sample of
almost $3\times 10^7$ $W$ pairs available at the 500~GeV ILC.   We
will also use the equality of the $Z$ left-right asymmetry $A_\ell$
for $e$  and $\tau$, since we will use a value of $A_\ell$
with contributions from $A_e$ and $A_\tau$.  (Dropping this equality
has only a minor  effect on our results.)  This assumption that can be tested through
meausurements of $\ee\to \ell^+\ell^-$ at 500~GeV.   A more 
complete Lagrangian would also include a four-fermion operator due to
new physics that
contributes directly to $G_F$.  However, this operator is already
constrained to have a $\Lambda$ scale above 8.5~TeV by LEP~2
data~\cite{RPP}, and this constraint will become much stronger when
data on $\ee\to\mu^+\mu^-$ at 500~GeV becomes available.   In our
discussion of the vector boson, lepton, and Higgs interactions, we
will 
also ignore leptonic terms that are mass-suppressed, including the
dimension-6 operators that correct the lepton-Higgs couplings and
lepton-gauge boson magnetic moment couplings.   The lepton terms that
correct the Higgs couplings will appear in Section 7. 
A more general analysis could incorporate more of these additional parameters and
the reactions that constrain them.

In this paper, we will avoid observables that involve the operators
similar to the last two lines of \leqn{firstL} that include quark
currents.  There is a very large number of these operators, two  for
each quark flavor.   Eventually, in Section 7, we will need to
consider these operators, but only in two specific linear
combinations.  When we refer to these operators later in the paper,
we will call the corresponding coefficients $c_X$, $c^\prime_{X}$.

Third, we will ignore the effects of the CP-violating operators in
\leqn{CPL}.   We will consider only CP-invariant observables, and so
the effects of these operators on our observables  will be of order $c_I^2$.  Actually, it
is possible to constrain the coefficients ${\tilde c}_{WB}$ and
${\tilde c}_{3W}$ 
below the percent level through the study of $\ee\to
W^+W^-$~\cite{ILCTDR} and to constrain ${\tilde c}_{WW}$ and
${\tilde c}_{BB}$ to the few-percent level through constraints from 
$h\to \gamma\gamma$ and $\ee\to Zh$.   We will present these latter
constraints in Sections 4 and 5.  At this level, these coefficients  would
give negligible contributions to our analysis.

\subsection{On-shell renormalization}

We then restrict ourselves to the SM Lagrangian plus the
perturbation \leqn{firstL}, considered in linear order.   Our analysis
of vector boson, lepton, and Higgs couplings then contains
14 parameters---the 4 SM parameters,
which we will take to be  $g$,
$g'$, $v$, and a Higgs coupling $\bar\lambda$, and the 10 parameters
in \leqn{firstL} (including $c_6$).   The dimension-6 operator coefficients alter the
SM expressions for precision electroweak observables and
thus shift the appropriate values for the Standard Model couplings.
In our analysis, we will deal with this by allowing the shifts of $g$,
$g'$, $v$ and $\bar\lambda$ from their SM values to be
free parameters in our fit.  In this
tree-level analysis, it is useful to think of $g$ and the other
couplings---and the parameters $s_w$ and $c_w$---as
 bare values set by fitting an expression that
includes the SM expectations and 
corrections perturbative in the $c_i$ to a set of measurements.  This
defines an on-shell renormalization procedure.

Using the notation
\beq 
    \delta A  =    {\Delta A\over A}  \ , 
\eeq{deltadefin}
we will write expressions for the deviation of  observables from their
SM predictions as linear combinations of the coefficients
$c_I$ and the deviations 
\leqn{deltadefin} of the SM parameters.
  A list of all of the expressions of this
type entering our fit is given in Appendix A.

Another approach to on-shell renormalization is given by the $S$, $T$
formalism \cite{PT}.   We will sketch the formulae for $S$, $T$
renormalization with EFT parameters in Appendix C.

The operators \leqn{firstL} also renormalize the kinetic terms of the
SM fields.    The contributions in \leqn{firstL} give
shifts of the SM kinetic terms
\beqa
   \L &=&  - {1\over 2} W_{\mu\nu}^+ W^{-\mu\nu} \cdot (1 - \delta Z_W)
   -  {1\over 4} Z_{\mu\nu} Z^{\mu\nu} \cdot (1 - \delta Z_Z)\CR
  & & \hskip 0.2in  -  {1\over 4} A_{\mu\nu} A^{\mu\nu} \cdot (1 - \delta Z_A)
   + {1\over 2} (\del_\mu h) (\del^\mu h) \cdot (1 - \delta Z_h) \ ,
\eeqa{Zfactors}
with 
\beqa
    \delta Z_W &=&  (8c_{WW}) \CR
    \delta Z_Z &=&   c_w^2 (8c_{WW})  + 2 s_w^2 (8c_{WB} ) +
      {s_w^4/c_w^2} (8 c_{BB}) \CR
    \delta Z_A &=&  s_w^2 \biggl(  (8c_{WW})  - 2 (8c_{WB} ) +
      (8c_{BB}) \biggr) \CR
    \delta Z_h &=&  - c_H  \ \ .
\eeqa{Zvals} 
We will rescale the boson fields to remove these factors from the
kinetic terms.  Then the $\delta Z$ factors will appear in the
vertices that we write below.  The field strength renormalization for
the Higgs field, proportional to $c_H$, plays a key role in
our analysis and in the general theory of Higgs
couplings~\cite{Han,Nate}.   It is important to note that the mass
eigenstates $Z$ and $A$ are not altered by the addition of
\leqn{firstL}.  The $c_T$ term shifts the mass of the $Z$ eigenstate
without mixing it with the $A$. However, \leqn{firstL} does induce  a
kinetic mixing between $Z$ and $A$, 
\beq
\Delta \L =   {1\over 2}\, \delta Z_{AZ} \,  A_{\mu\nu} Z^{\mu\nu}  
 \  ,
\eeq{AZmix}
with 
\beq
  \delta Z_{AZ} =  s_w c_w  \biggl(   (8c_{WW})  - (1 - {s_w^2\over
    c_w^2})(8c_{WB})  - {s_w^2\over c_w^2} (8c_{BB}) \biggr) \  .
\eeqn
We will treat this effect in perturbation theory.

The masses of the bosons are then given by 
\beqa
 m_W^2 &=&   {g^2 v^2 \over 4} \bigl(1 + \delta Z_W\bigr) \CR
m_Z^2 &=&   {(g^2+g^{\prime 2}) v^2 \over 4} \bigl(1  - c_T + \delta
Z_Z \bigr) \CR
m_h^2 &= &  2 \bar \lambda v^2 (1 + \delta Z_h)
\eeqa{SMmasses}
where
\beq
   \bar \lambda =  \lambda (1 + \thalf c_6) \ . 
\eeqn
It is useful to take $\bar \lambda$ as a basic coupling, since the Higgs quartic
coupling $\lambda$ and the dimension-6 coefficient $c_6$ appear only
in this combination until we actually encounter the triple Higgs coupling
in our analysis.   The formulae \leqn{SMmasses} are not precise for
the absolute values of the masses without inclusion of loop
corrections.  However, the differential relations
\beqa
 \delta m_W &=&   \delta g + \delta v + \half \delta Z_W\CR
\delta m_Z  &=&   c_w^2 \delta g + s_w^2 \delta g' + \delta v   -
\half c_T + \half \delta Z_Z \CR
\delta m_h  &= &  \half \delta \bar\lambda + \delta v + \half \delta Z_h
\eeqa{SMmassdiffs}
are accurate for small deviations.

To expand other precision electroweak observables, it is useful to
expand expressions built from the bare couplings
\beqa
\delta s_w &=& - c_w^2 ( \delta g - \delta g')\CR
\delta c_w &=& s_w^2 (\delta g - \delta g')
\eeqan
The physical electric charge is expanded as
\beq
\delta e = \delta (4\pi \alpha(m_Z^2))^{1/2} =  s_w^2 \delta g + c_w^2
\delta g' +\half \delta Z_A \ .
\eeq{edefine}
The Fermi constant obtains a contribution from one of the Higgs-lepton
current-current operators. It also received contributions $(1 + \delta
Z_W)$ from the $W$ mass and coupling that cancel between numerator and
denominator.   Then
\beq
 \delta G_F  = 1 - 2 \delta v + 2 c_{HL}^\prime .
\eeqn

\begin{figure}
\begin{center}
\includegraphics[width=0.70\hsize]{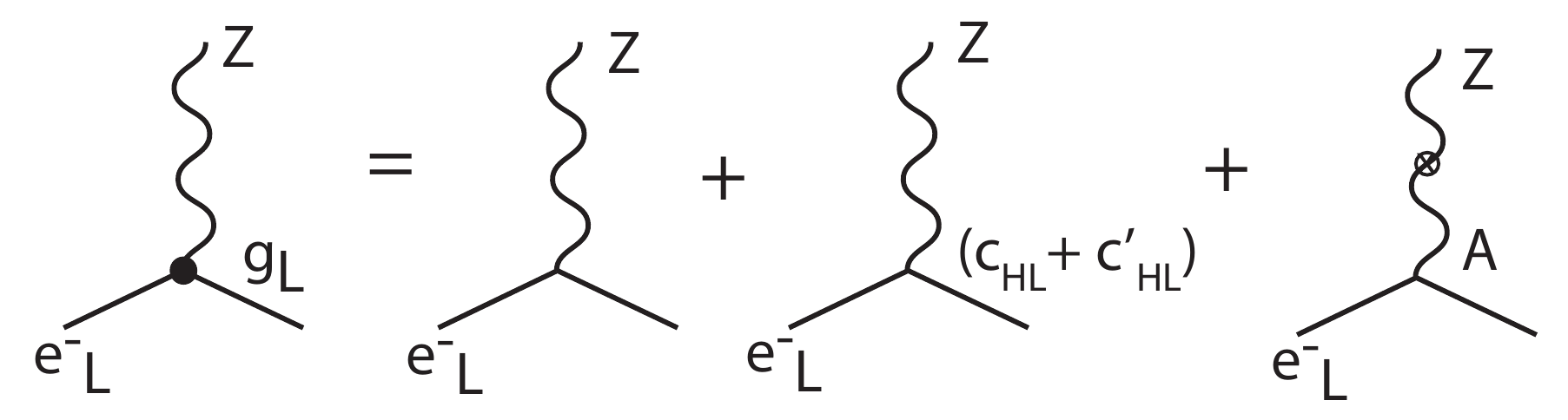}
\end{center}
\caption{Contributions to $g_L$,  the left-handed electron
  coupling to the $Z$, including the effects of contact interactions
  and $AZ$ kinetic mixing.   The contributions to $g_R$ have a similar structure.}
\label{fig:Zcouplings}
\end{figure}

In writing the $Z$ boson couplings of the light leptons, it is
convenient to include the contribution due to the $AZ$ kinetic mixing
in \leqn{AZmix}, as shown in Fig.~\ref{fig:Zcouplings}. Then the left-
and right-handed charged lepton couplings are
\beqa
g_L &=& {g\over c_w}\biggl[ (-\half + s_w^2)(1 + \half \delta Z_Z)
        - \half (c_{HL} + c_{HL}^\prime)  - s_wc_w \delta Z_{AZ}
        \biggr] \CR
g_R &=& {g\over c_w}\biggl[ ( + s_w^2)(1 + \half \delta Z_Z)
        - \half c_{HE}  - s_wc_w \delta Z_{AZ}
        \biggr] 
\eeqa{gLgR}
The $W$ coupling to leptons is given by 
\beq
g_W =  g \ ( 1 + c^\prime_{HL} + \half \delta Z_W) \ \ . 
\eeqn
In Section 2.5, we will introduce $Z$ coupling to $W^+W^-$.  Its value
is
\beq
g_Z = g c_w \ (1 + \half \delta Z_Z + {s_w\over c_w} \delta Z_{AZ} ) \
. 
\eeqn
The differentials of these expressions are written in Appendix A.

\subsection{Precision electroweak constraints}

The five parameters $m_W$, $m_Z$, $m_h$, $\alpha(m_Z)$, $G_F$
constrain independent combinations of the four Standard Model couplings and the
dimension-6 coefficients.  Most of the power of the precision
electroweak constraints on our parameter set is given by adding two
further, very precise, measurements from $Z$ physics.   We choose
these to be $\Gamma_\ell$, the partial width of the $Z$
to a lepton, and $A_\ell$, the left-right asymmetry of the $Z$
coupling to leptons.   All dimension-6 corrections to these
coefficients are already incorporated into $g_L$ and $g_R$,
 so the differentials of these parameters are given
in terms of \leqn{gLgR} by
\beqa
  \delta \Gamma_\ell &=& \delta m_Z +  2{ g_L^2 \delta g_L + g_R^2
    \delta g_R \over g_L^2 + g_R^2} \CR
\delta A_\ell &=&  {4 g_L^2 g_R^2( \delta g_L - \delta g_R)\over
  g_L^4-g_R^4 } 
\eeqan
Note that no dimension-6 operators involving quarks enter an analysis
based on these observables.

The values that we will here use for the precision electroweak observables,
and their errors, are the current values, from \cite{RPPEW}.   For the
$A_\ell$, we take the value corresponding to the average of
$\sin^2\theta^{lept}_{eff} $ presented in Section 7.3.4 of~\cite{LEPEWWG}.  
 These values are shown in
Table~\ref{tab:PEW}.  

For the analysis in Section 7, we will need to make use of the
measurements of the total width of the $Z$ and $W$.  So we include
those current values also in Table~\ref{tab:PEW}.  

Our analysis will benefit from improvements in some of the precision
electroweak parameters that we expect to see in the era of $\ee$
experiments.
 The uncertainties on $m_W$ 
measurements are expected to be improved to 5~MeV already at
LHC~\cite{SnowmassEW}.  The ILC is expected to improve the error on
the Higgs boson mass to 15~MeV by recoil mass fitting of $\ee\to Zh$
events in which the $Z$ decays to leptons~\cite{Recoil1}. It is not so
easy
 to obtain a very precise direct measurement of the
$W$ total width.  Today, this width is known only to 2\% accuracy.
However, with the use of constraints from other precision electroweak
observables
and measurements of $\ee\to W^+W^-$, our EFT formalism predicts the 
partial width $\Gamma(W\to \ell\nu)$ to an accuracy of 0.06\%.
  Using the large statistics available at the
ILC---$3\times 10^7$ pairs---it will be possible to apply a tag and probe
method to make a very
precise measurement of the branching ratio $BR(W\to\ell\nu)$.  We then
expect that the total width $\Gamma_W$ can be known to better than
0.1\%.  Running an $\ee$ collider at  the $Z$ resonance to create at least
$10^9$ $Z$ bosons would be expected to improve the errors on $A_\ell$ and
$\Gamma_\ell$ by an order of magnitude~\cite{Monig}. However, we will not make use
of that possibility in the estimates given in this paper.

\begin{table}
\begin{center}
\begin{tabular} {lcccc}
Observable  &    current value &   current $\sigma$ &  future
$\sigma$ &  
 SM best fit value \\ \hline
$\alpha^{-1}(m_Z^2) $&        128.9220            &  0.0178       &  
&     (same)             \\ 
$G_F$   ($10^{-10}$ GeV$^{-2}$)    &     1166378.7        & 0.6        &
&       (same)                           \\
$m_W$  (MeV) &          80385                  &  15   &   5    &
80361  \\
$m_Z$  (MeV)  &          91187.6                  &   2.1   & &
91188.0 \\ 
$m_h$   (MeV) &      125090    &   240    &   15    &   125110   \\ 
$A_\ell $ &         0.14696      &      0.0013                   &
&          0.147937  \\          
$\Gamma_\ell$  (MeV)   &         83.984                   &   0.086
&  
&        83.995                            \\
$\Gamma_Z$ (MeV)  &         2495.2                   &   2.3   &
&              2494.3                      \\
$\Gamma_W$  (MeV) &        2085                     &    42     &
2  &     2088.8 \\
\end{tabular}
\end{center}
\caption{Values and uncertainties for precision electroweak
  observables used in this paper.   The values are taken from
  \cite{RPPEW}, except for the averaged value of $A_\ell$, which
  corresponds to the averaged value of 
  $\sin^2\theta_{eff}$ in  \cite{LEPEWWG}.  The best
  fit values are those of the fit in \cite{RPPEW}.  For the
  purpose of fitting Higgs boson couplings as described in Section 7,
  we use improvements in some of the errors expected from
  LHC~\cite{SnowmassEW} and ILC~\cite{Recoil1}.  The improved
  estimate of the $W$ width is obtained from  $\Gamma_W = \Gamma(W\to
  \ell \nu)/BR(W\to \ell \nu)$. }
\label{tab:PEW}
\end{table}

\subsection{$W$, $Z$,  and Higgs boson vertices}

Starting from the Lagrangian \leqn{firstL} and using the prescriptions
in Section 2.3, we can work out expressions for the various coupling
constants that appear in the $W$ and Higgs interactions.

\begin{figure}
\begin{center}
\includegraphics[width=0.70\hsize]{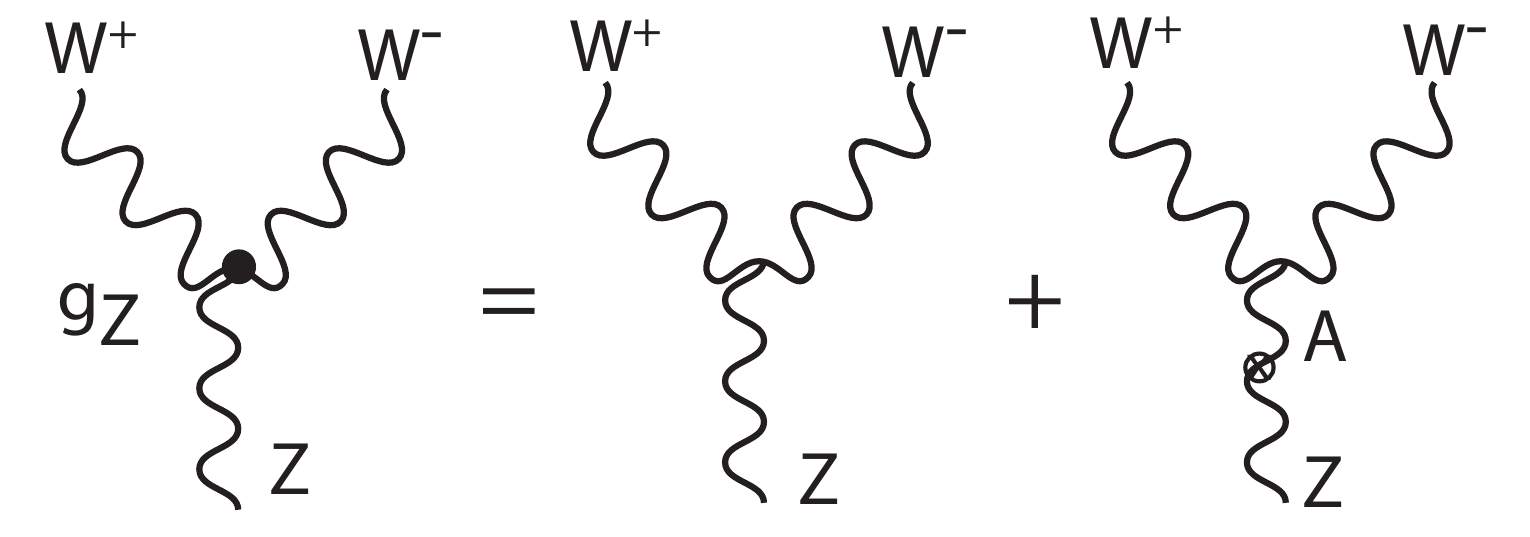}
\end{center}
\caption{Contributions to $g_Z$,  the coupling of the $Z$ boson to
  $W^+W^-$, including in particular the effect of $AZ$ kinetic mixing.}
\label{fig:ZWcoupling}
\end{figure}

The three-boson vertices involving the $W$ boson are canonically
written~\cite{Hagiwara}
\beqa
\Delta \L_{TGC} &=& i g_V \biggl\{    V^\mu \bigl( { \hat W}^-_{\mu\nu}
W^{+\nu} -  {\hat W}^+_{\mu\nu}
W^{-\nu} \bigr)  +  \kappa_V  W^+_\mu W^-_\nu {\hat V}^{\mu\nu} \CR
& & \hskip 0.7in + {\lambda_V \over m_W^2}{\hat W}^-_\mu{}^\rho {\hat W}^+_{\rho \nu}
 {\hat  V}^{\mu\nu}  \biggr\} \ ,
\eeqa{TGC}
where $V = A$ or $Z$ and 
\beq 
     {\hat V}_{\mu\nu} = \del_\mu V_\nu - \del_\nu V_\mu
\eeqn
is the linear part (only) of the field strength tensor.  Note that we
have absorbed the constant in front of the first term in \leqn{TGC}
into the overall coupling $g_A$ or $g_Z$.   Then this 
formula has 6 parameters.  Of these $g_A$ must turn out to equal the physical
electron charge $e$ in \leqn{edefine}, since this is also the charge
of the $W$.   It is a simple exercise to verify this explicitly.
We define the charge $g_Z$ to include the effect of $AZ$ kinetic
mixing, as shown in Fig.~\ref{fig:ZWcoupling}.  Then  the charge $g_Z$ is given by
\beq
    g_Z =  g c_w (1 + \half \delta Z_Z +   {s_w\over c_w} \delta
    Z_{AZ} )
\eeqn
The remaining parameters are given by 
\beqa
  \kappa_A    & =&  1 + (8 c_{WB} )\CR
 \kappa_Z     &=&  1 -  {s_w^2\over c_w^2}\,(8 c_{WB} )\CR
\lambda_A &=& \lambda_Z = - 6 g^2 c_{3W} 
\eeqan
Because of the relations between these expressions, the measurement of
the $WWA$ and $WWZ$ vertices contribute three (not five) additional constraints on
our 14 variables.  We will work out the form of these constraints in
Section 3.

In a similar way, we will write the Lagrangian for  the Higgs boson and its
coupling to vector bosons in a canonical form as
\beqa
\Delta \L_{h} &=&\half \del_\mu h \del^\mu h  - \half m_h^2 h^2 - (1 +
\eta_h) \bar \lambda v h^3 
 + {\theta_h \over v} h \del_\mu h \del^\mu h \CR
& &\hskip 0.3in +(1+\eta_W) {2 m_W^2\over v} W^+_\mu W^{-\mu} h  +(1 +\eta_{WW}) {m_W^2\over
  v^2} W^+_\mu W^{-\mu} h^2\CR
& & \hskip 0.3in +(1 + \eta_Z)  {m_Z^2\over v} Z_\mu Z^\mu h  +\half (1+\eta_{ZZ}) {m_Z^2\over
  v^2} Z_\mu Z^\mu h^2\CR
& &\hskip 0.2in +  \zeta_{W} {\hat W}^+_{\mu\nu}{\hat  W}^{-\mu\nu} \biggl( {h\over v} +
\half {h^2\over v^2}\biggr) +\half \zeta_{Z} {\hat Z}_{\mu\nu}{\hat  Z}^{\mu\nu}
\biggl( {h\over v} +
\half {h^2\over v^2}\biggr) \CR 
& &\hskip 0.2in + \half \zeta_{A} {\hat A}_{\mu\nu}{\hat  A}^{\mu\nu} \biggl( {h\over v} +
\half {h^2\over v^2}\biggr)  + \zeta_{AZ} {\hat A}_{\mu\nu}{\hat  Z}^{\mu\nu} \biggl( {h\over v} +
\half {h^2\over v^2}\biggr) \ .
\eeqa{Higgsform}
The 6 parameters in the first two lines of this  equation are given to first order in the EFT
coefficients by 
\beqa
\eta_h &=&  \delta \bar  \lambda + \delta v  - \thalf c_H + c_6\CR
  \theta_h &=&  c_H \CR
\eta_W &=&  2\delta m_W - \delta v  - \half c_H \CR
\eta_{WW} &=& 2\delta m_W -2 \delta v  - c_H  \CR
\eta_Z &=&  2\delta m_Z - \delta v - \half c_H - c_T \CR
\eta_{ZZ} &=&  2\delta m_Z - 2\delta v - c_H - 5 c_T \  .
\eeqa{HiggsEFTforc}
The four parameters in the last two lines are given by 
\beqa
    \zeta_W= \delta Z_W &=&   ( 8c_{WW})\CR
   \zeta_Z = \delta Z_Z &=&   c_w^2 (8c_{WW} )+ 2 s_w^2 (8c_{WB}) + {s_w^4\over c_w^2} 
 (8 c_{BB}) \CR
\zeta_A = \delta Z_A &=& 8 s_w^2 \biggl( (8c_{WW} ) - 2(8c_{WB}) + (8c_{BB})\biggr)  \CR
 \zeta_{AZ} = \delta Z_{AZ}  &=&  s_w c_w \biggl( (8c_{WW}) -  (1 - {s_w^2\over
   c_w^2} ) (8c_{WB}) - {s_w^2\over c_w^2}(8 c_{BB}) \biggr)\  .
\eeqa{zetaforc}

It is important to note that \leqn{Higgsform} contains a second structure for the triple
Higgs vertex, with coefficient $\theta_h$.  In double Higgs
production,
 this term cannot be separated
from the Standard Model triple Higgs  coupling except by high
statistics measurement of the $m(hh)$ distribution.
 In our analysis, this contribution will be fixed by the
 determination of $c_H$ through precision measurement of single
Higgs production.

The Lagrangian \leqn{firstL} also contains contact interactions
between the $Z$, Higgs, and lepton fields,
\beqa
    \Delta \L_{eehZ} &=&  -{g\over 2c_w} (c_{HL}-c^\prime_{HL}) (\bar \nu_L
    \gamma_\mu \nu_L) Z^\mu  (1+2 {h\over v}+ {h^2\over v^2}) \CR
& &  -{g\over 2c_w} (c_{HL}+c^\prime_{HL}) (\bar e_L
    \gamma_\mu e_L) Z^\mu  (1+2 {h\over v}+ {h^2\over v^2}) \CR
 & &  -{g\over 2c_w} (c_{HE}) (\bar e_R
    \gamma_\mu e_R) Z^\mu  (1+2 {h\over v}+ {h^2\over v^2}) \CR
& &  {g\over \sqrt{2}} (c^\prime_{HL}) (\bar e_L
    \gamma_\mu \nu_L \ W^{-\mu} +\bar \nu_L
    \gamma_\mu  e_L \ W^{+\mu} (1+2 {h\over v}+ {h^2\over v^2}) \ .
\eeqa{Higgscontact}
We will discuss the use of \leqn{Higgsform} and \leqn{Higgscontact} to construct expressions for 
the cross sections for $\ee\to Zh$ and $\ee\to Zhh$ in Sections~5 and
6.    Effects of the contact interactions in \leqn{Higgscontact}
  on these reactions have previously been studied in \cite{Shaouly},
  where it was pointed out that the coefficients typically appear in
  the form $c_i\cdot s/m_Z^2$.

\section{Constraints from $\ee\to W^+W^-$}

We now discuss the constraints on the dimension-6 coefficients
coming from measurements of $\ee\to W^+W^-$.  The constraints coming
from the LEP and LHC experiments have been discussed already in
\cite{Riva,Rivatwo}.
However, future $\ee$ experiments will have additional advantages.
Since $\ee\to W^+W^-$ is the process with the largest cross section in
high-energy $\ee$ annihilation, very high statistics will be available.
 By making use especially of the mode in which one $W$
decays hadronically and one decays leptonically, the full kinematics
of the $W^+W^-$ production and decay can be reconstructed for each
event.   Changing the beam polarization from $e^-_Le^+_R$ to
$e^-_Re^+_L$ is an order-1 effect.  Using all of these handles, it is
possible to probe very accurately for the effects of modifications of
the Standard Model.

\begin{figure}
\begin{center}
\includegraphics[width=0.70\hsize]{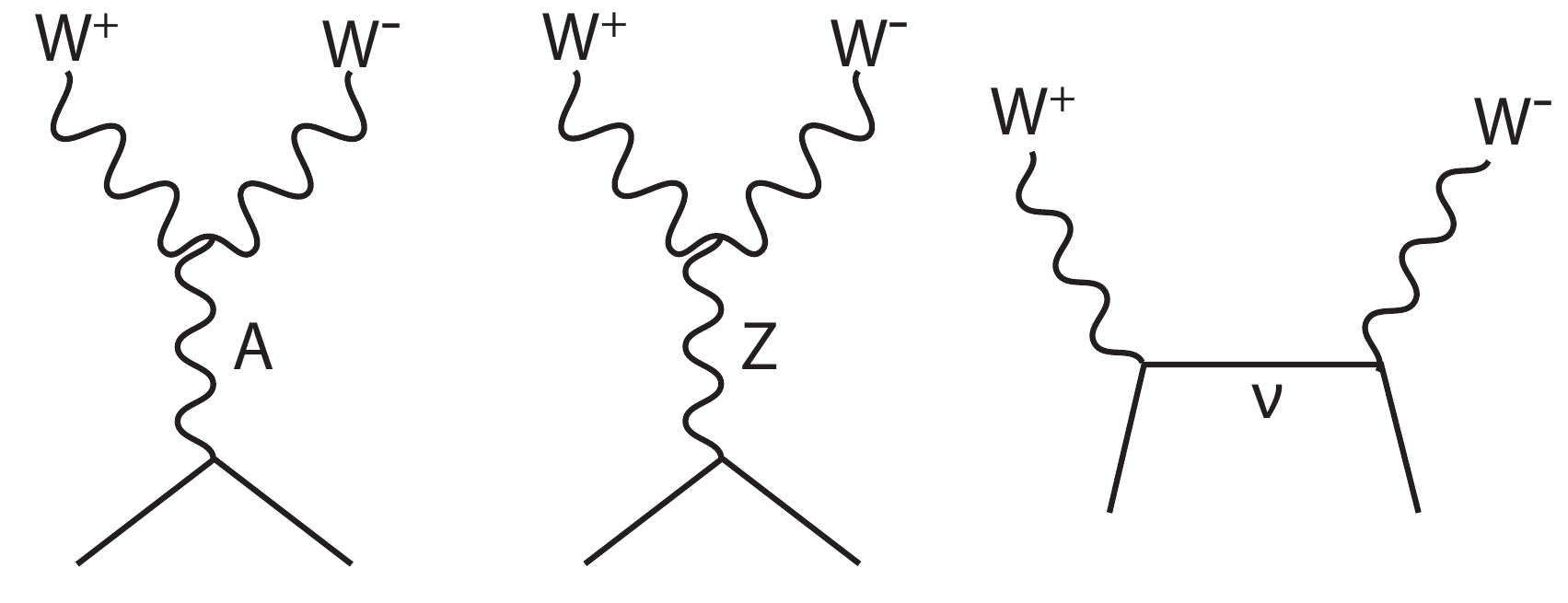}
\end{center}
\caption{Feynman diagrams contributing to the amplitudes for $\ee\to W^+W^-$.}
\label{fig:eeWW}
\end{figure}

At the tree level, the amplitudes for $\ee\to W^+W^-$ are derived from
the diagrams shown in Fig.~\ref{fig:eeWW}.   Typically, the
predictions of these diagrams (with higher-order electroweak
corrections) are compared to 
data by assuming that the vertices between leptons and gauge bosons
have exactly the SM form while the $WWA$ and $WWZ$ vertices can
contain additional terms induced by new physics.   In the EFT, there
are relations within the full set of phenomenological parameters
present in \leqn{TGC}.   These relations, which follow from the
$SU(2)\times  U(1)$ gauge invariance of the full
theory~\cite{WWrelations},   are written in our notation as  $g_A = e$ and 
\beq  
   (\kappa_Z -1 ) = - {s_w^2\over c_w^2} (\kappa_A -1) \qquad
   \lambda_Z = \lambda_A \ , 
\eeqn
Then the three parameters $g_Z$, $\kappa_A$, $\lambda_A$ are
extracted from the data.   The most incisive projections of the
capabilities of future $\ee$ collliders to extract these parameters
were done using this method by  Marchesini~\cite{Marchesini} and 
Rosca~\cite{Rosca}.  These studies used full simulation with the ILD 
detector model at the ILC.   The precise uncertainties expected at 500~GeV
with the expected  4~ab$^{-1}$ data sample, including
their correlations, are given in Appendix B. 

The assumption that the SM lepton-gauge boson vertices are unmodified
is justified to a certain extent by the strength of the precision
electroweak constraints on those vertices, but it is not completely
consistent as an expansion in the dimension-6 EFT coefficients. This
point is made explicitly by Falkowski and Riva in \cite{Riva},
although they neglect this effect in their analysis for the practical
reason that it is unimportant in fitting LEP and LHC data.  For
higher-accuracy measurements, one should in principle
refit the experiment data with a
formula that includes all of the terms linear in dimension-6
coefficients  fully.   Here we propose a simplified treatment
of this issue.  It is well appreciated that the process $\ee\to
W^+W^-$ is especially sensitive to new physics corrections because 
the helicity amplitudes for this process contain terms proportional to
the $c_I$ coefficients enhanced by $s/m_W^2$.    So, we will propose
definitions of effective values of $g_Z$, $\kappa_Z$, and $\lambda_Z$
that agree with the standard definitions when the precision
electroweak constraints are exact and otherwise include the
deviations from SM precision electroweak  proportional to  $c_I s/m_W^2$.

To do this, we compute the high-energy limit of the helicity
amplitudes for $\ee\to W^+W^-$ for the case in which both $W$ bosons
have longitudinal polarization.   For both possible  beam polarization
states, the results have the form
\beq 
i{\cal M} \to - i \sin\theta {s\over 2m_W^2} {\cal A}_{L,R}\ . 
\eeqn
 For $e^-_Re^+_L$, the neutrino
diagram does not contribute and we find
\beq 
{\cal A}_R =  e^2 \kappa_A + g_R
g_Z \kappa_Z  \ . 
\eeq{calAR}
For $e^-_Le^+_R$, all three diagrams contribute and we find
\beq 
{\cal A}_L =  e^2 \kappa_A + g_L
g_Z \kappa_Z  - { g_W^2 \over 2} \ . 
\eeq{calAL}
It is easy to check that both quantities \leqn{calAR}, \leqn{calAL}  vanish when the
coupling constants take their SM values (including $\kappa_A =
\kappa_Z = 1$).  Note that both amplitudes are independent of
$\lambda_A$ and $\lambda_Z$.  The $\lambda$  parameters multiply a different
$s/m_W^2$ term that appears in the helicity amplitudes for the
production of two transversely polarized $W$ bosons.

We propose, then, to use the following quantities to express the
constraints on the $c_I$ from $\ee\to W^+W^-$:
\beqa
(g_{Z,eff} -1) &=&  {1\over g^2c_w^2  } \biggl(  2 \Delta {\cal A}_L -
\Delta{\cal A}_R \biggr)\CR
(\kappa_{A,eff} -1) &=&  {1\over g^2  } \biggl( 2 \Delta {\cal A}_L +
{c_w^2 - s_w^2\over s_w^2} \Delta{\cal A}_R \biggr) \CR
\lambda_{A,eff} &=& \lambda_A
\eeqa{effectiveWW}
The right-hand sides of \leqn{effectiveWW} can be expanded in terms of
the variations of SM parameters and the $c_I$.   The expansions are
written out in Appendix A.  These formulae can be considered to be the
quantities constrained by the analyses of \cite{Marchesini} and
\cite{Rosca}.
The measurements of $W$ vertices  at LHC should be compared to similar formulae that
involve the $c_{HX}$ parameters for the various quark species that
participate in the observed processes.

\section{Constraints from $h\to \gamma\gamma$ and $h\to \gamma Z$}

We have now explained how to constrain 10 combinations of the 14
parameters in our analysis.  Before we come to precision Higgs physics
in $\ee$ collisions, there is one more important constraint that we
can apply.

The Higgs boson decays $h\to\gamma\gamma$ and $h\to Z\gamma$ receive
their first SM contributions at the one-loop level.   In both cases,
these contributions are very small.   However, both decays receive
tree-level contributions from the dimension-6 Lagrangian, proportional
to the coefficients $\zeta_{A}$ and $\zeta_{AZ}$ in \leqn{Higgsform}.
If these decays are observed to have rates close to their SM values,
the parameters $\zeta_A$ and $\zeta_{AZ}$ are constrained to have
values that are small fractions of the already suppressed SM decay
amplitudes~\cite{Disclaimer}.  Already, the constraints from LHC on $h\to \gamma\gamma$
are quite strong.  Eventually, the measurement of these modes
 will provide an extremely strong constraint on the
parameter $c_{BB}$ and a significant constraint on the parameter
$c_{WB}$. 

We now analyze  this point in more detail.   The $h\to\gamma\gamma$
decay amplitude has the form 
\beq
  i{\cal M} =  i {\cal A}  (q_1^\mu \cdot q_2^\nu - q_1^\nu q_2^\mu)
  \eps^*_{1\mu} \eps^*_{2\nu} \ . 
\eeqn
The $\zeta_A$ term contributes an extra factor
\beq
\Delta {\cal A} =  { 2 \zeta_A\over v} \ . 
\eeqn
Then
\beqa
    \delta \Gamma(h\to \gamma\gamma) &=&   4 {\zeta_A\over v} \biggl[
 {   m_h^3 \over 64\pi \Gamma(h\to \gamma\gamma)|_{SM}} \biggr]^{1/2}\CR
 & = &  526 \ \zeta_A \ . 
\eeqa{fromzetaA}
In a similar way, we find
 \beqa
    \delta \Gamma(h\to Z\gamma) &=&   4 {\zeta_{AZ}\over v} \biggl[
  {  m_h^3 (1 - m_Z^2/m_h^2)^3\over 32\pi \Gamma(h\to Z\gamma)|_{SM}} \biggr]^{1/2}\CR
 & = &  290 \ \zeta_{AZ} \ . 
\eeqa{fromzetaAZ}
We must add to these expressions the variation of the SM predictions
for the decay widths with respect to the SM parameters.   The complete
expressions are given in Appendix A.   We omit loop-suppressed
corrections from the $c_I$ coefficients.   In fact, \leqn{fromzetaA}
and \leqn{fromzetaAZ} are by far the dominant effects.

It is not possible to measure absolute Higgs decay widths at
the LHC, because there is no strategy to obtain the total Higgs width
to high accuracy.  But, it is possible to measure ratios of branching ratios
from which the total Higgs width cancels out.  Since the measurement
of each Higgs boson final state at the LHC requires its own strategy,
measurements of ratios of branching ratios are typically limited by
the separate systematic errors from production and detection of the
two processes that are compared.   The only exceptions of which we are
aware are the ratios
\beq
    {BR(h\to ZZ^*\to 4\ell)\over BR(h\to \gamma\gamma)} \  , \quad 
  {BR(h\to Z\gamma)\over BR(h\to \gamma\gamma)} \  , \quad  
 {BR(h\to \mu^+\mu^-)\over BR(h\to \gamma\gamma)} \  . 
\eeqn
These ratios all involve rare decay modes in which the Higgs can be reconstructed
as a resonance, so they can be detected in the major production mode
$gg\to h$ at low Higgs boson $p_T$.   The ATLAS Collaboration has
estimated that first of these ratios can be measured to 3.6\% accuracy
in the full LHC program with 3000~fb$^{-1}$~\cite{ATLASfuture}.    We believe that, with
an analysis specifically designed to cancel systematic errors, it will
be possible to reach the statistics-limited accuracy of 2\%.   This
means that the combination of $c_I$ coefficients in $\zeta_A$ will be
constrained to   $10^{-4}$ accuracy.  For
the more difficult decays to $Z\gamma$ and $\mu^+\mu^-$, ATLAS has estimated eventual
accuracies  in these ratios of   31\% and 12\%,
respectively~\cite{ATLASfuture,ATLASZgam}.   These measurements can
be converted to partial width measurements when the absolute partial
width $\Gamma(h\to ZZ^*) $ is measured at future $\ee$ colliders.

A CP-violating contribution to the $h\to \gamma\gamma$ decay from the
operators in \leqn{CPL} would give a strictly additive contribution to the total rate of
$h\to \gamma\gamma$ decay.   A constraint of 2\% on deviations from
the SM in $\Gamma(h\to
\gamma\gamma)$ would then place a constraint on the ${\tilde c}_{BB}$
coefficient in \leqn{CPL} at about  1\%.  This is a strong enough
constraint that this CP-violating coefficient can be ignored in our analysis.

\section{Constraints from $\ee\to Zh$}

At this point, all of the original 13 parameters are strongly
constrained except for $c_{WW}$ and the parameters $c_H$ that
appears only in Higgs decays.  In this section, we
explain how to determine them through the study of the process $\ee\to
Zh$.    Our results for the total cross section in $\ee\to Zh$ are
similar to those in \cite{Maxim}, but we also consider other
observables of this process.

It is important to recall here that the parameter $c_H$ appears in the
normalization of all Higgs couplings thorugh the field strength
renormalization
\leqn{Zvals}.  Thus, it is not possible to
determine $c_H$ unambiguously  without measuring an absolute Higgs production or
decay rate.  Measurements of $\sigma\cdot BR$ are not sufficient.
The total cross section for $\ee\to Zh$ can be measured by tagging a
$Z$ boson at the correct energy to be recoiling against a Higgs boson
without the need for any information from the Higgs decay products.
Thus, in principle, it provides a way to measure $c_H$.  In the EFT
formalism, there are complications from the fact that other EFT
parameters also affect the size of the cross section.  We will discuss
the untangling of  this parameter dependence at the end of this
section and again in Section 8. 

The amplitudes for the reaction $\ee\to Zh$ have a very simple form.
For each initial polarization state $e^-_Le^+_R$ or $e^-_Re^+_L$,
there are three helicity amplitudes, corresponding to the three
possible $Z$ boson polarization states.  However, the two amplitudes
with transverse $Z$ polarizations are related by CP, so there are only
two independent amplitudes.  Further, at the tree level within the EFT
description, at a fixed CM energy, these amplitudes can be written
with only two 
free parameters.

To describe these amplitudes, it is most convenient to begin by
considering only the contribution from 
$s$-channel $Z$ boson exchange using the 
very simple---apparently, oversimplified---phenomenological
Lagrangian
\beq
\Delta \L =  {m_Z^2\over v} (1+ a)\, h \, Z_\mu Z^\mu + \half {b\over v}\, h\,
Z_{\mu\nu} Z^{\mu\nu} \ . 
\eeqn
Let $E_Z$ and $k$ be the energy and momentum of the $Z$ in the CM
frame.  Then we find, 
for $e^-_Le^+_R$, this gives the helicity amplitudes
\beqa
  i\M (\ee \to Z(\pm 1) h)  &=&  i g_L {\sqrt{2s}\over (s-m_Z^2)}
  \biggl[ (1+a)  {m_Z^2
    \over v} + b  {E_Z \sqrt{s}\over v} \biggr] (\cos\theta \pm
    1) \CR
  i\M (\ee \to Z(0) h)  &=&  i g_L {\sqrt{2s}\over (s-m_Z^2)}
  \biggl[ (1+a)  {m_Z E_Z
    \over v} + b  {m_Z \sqrt{s}\over v} \biggr] (\sqrt{2} \sin\theta) \ ,
\eeqa{Zhhelicity}
where the $\theta$ is the polar angle in production and the amplitudes
are labelled by the $Z$ helicity.
For $e^-_Re^+_L$, the helicity amplitudes take the same form except
for the substitution of $g_R$ for $g_L$ and $(\cos\theta \mp 1)$ for 
$(\cos\theta\pm 1)$. 
These helicity amplitudes control the total cross section, the $Zh$
angular distributions, and the distributions of the $Z$ decay angle.
In particular, the total cross section for a polarized initial state
is given for $e^-_Le^+_R$ by 
\beqa
 \sigma(e^-_Le^+_R \to Zh) & =& {1\over 6 \pi}{g^2\over  c_w^2}
 {m_Z^4/v^2\over (s-m_Z^2)^2}\cdot  {2k\over \sqrt{s}} \cdot \bigl(2 + {E_Z^2\over
   m_Z^2}\bigr)\CR
 & & \cdot (\half - s_w^2)^2 \biggl( (1+ 2a ) + 6 b {E_Z\sqrt{s}\over
   m_Z^2 (2 + E_Z^2 /m_Z^2)} \biggr)  \ ,
\eeqan
For $e^-_Re^+_L$, we have the same expression 
with the substitution $(\half - s_w)^2 \to s_w^2$.

In \cite{Ogawa}, it is shown how to obtain the values of the
parameters $a$ and $b$ by fitting to the production and decay angular
distributions in $\ee\to Zh$ events.   Using full simulation with the
ILD detector model and the 4 ab$^{-1}$ event sample expected for the
ILC at 500 GeV,  it is shown that the parameters $a$ and $b$ can be constrained
at  the percent level.   The precise uncertainties expected, including
their correlation, are given in Appendix B.   To the accuracy of the
study, these uncertainties are independent of the initial $\ee$
polarization state. 

\begin{figure}
\begin{center}
\includegraphics[width=0.70\hsize]{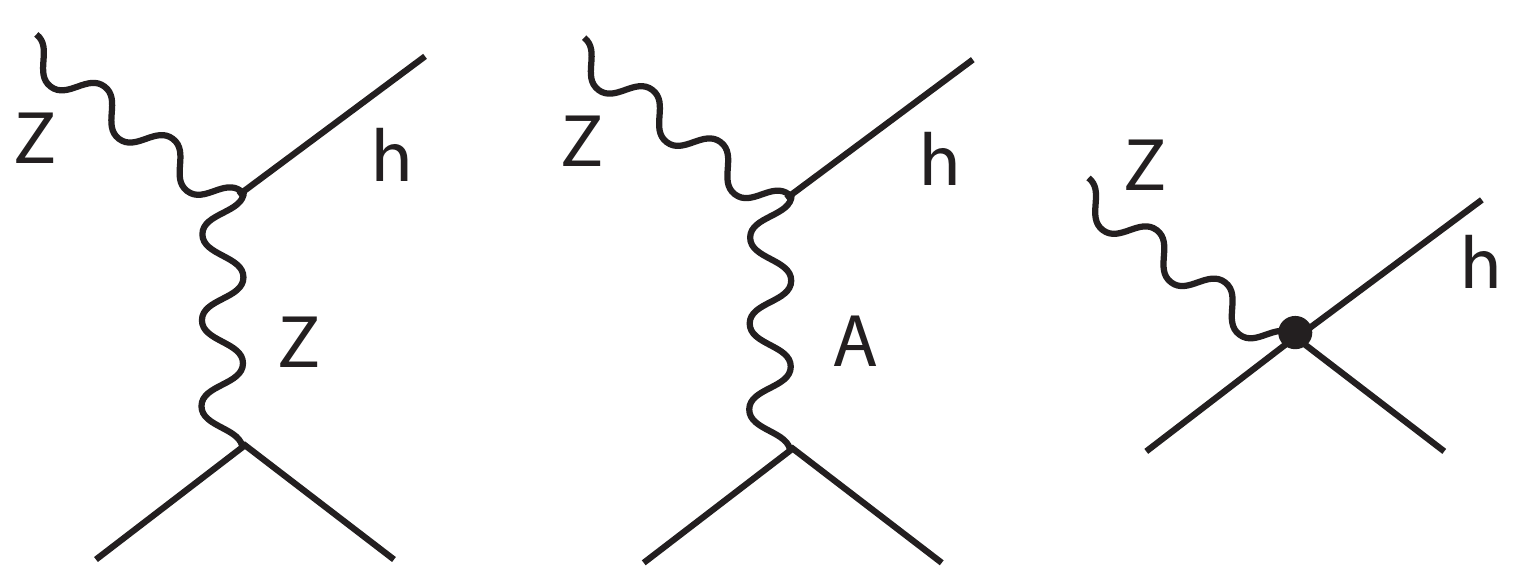}
\end{center}
\caption{Feynman diagrams contributing to the amplitudes for $\ee\to Zh$.}
\label{fig:Zhdiagrams}
\end{figure}

We can connect this analysis to the EFT parametrization of new physics
effects by noting that the complete tree-level calcuation of the
helicity amplitudes for $\ee\to Zh$ gives results that are still of
the form of \leqn{Zhhelicity} for appropriate identification of the
parameters $a$ and $b$.    The complete set of Feynman diagrams is shown in
Fig.~\ref{fig:Zhdiagrams}.     This includes a diagram with
$s$-channel $Z$ exchange (with the $s$-channel $AZ$ mixing already
included in the expressions for $g_L$ and $g_R$), a diagram with
$s$-channel photon exchange that makes use of the $\zeta_{AZ}$ vertex,
and a contact interaction proportional to $(c_{HL} + c^\prime_{HL})$
  or $c_{HE}$.    Diagrams with $AZ$ kinetic mixing on the
  final-state line are of order $c_I^2$ and so are not included in our
  calculation.

Evaluating the diagrams in Fig.~\ref{fig:Zhdiagrams}
and also expanding the SM dependence of the prefactors, we find,
for $e^-_Le^+_R$,
\beqa
 a_L &=&  \delta g_L + 2 \delta m_Z - \delta v + \eta_Z +
 {(s-m_Z^2)\over 2 m_Z^2 (1/2 - s_w^2)} (c_{HL} + c^\prime_{HL})  +
 k_Z \delta m_Z + k_h \delta m_h \CR
b_L &=&  \zeta_Z + {s_w c_w\over (1/2 - s_w^2)} {(s - m_Z^2 )\over s}
\zeta_{AZ} \ .
\eeqa{abLforms}
Similarly, for $e^-_Re^+_L$,
\beqa
 a_R &=&  \delta g_R + 2 \delta m_Z - \delta v + \eta_Z -
 {(s-m_Z^2)\over 2 m_Z^2( s_w^2)} c_{HE}  +
 k_Z \delta m_Z + k_h \delta m_h \CR
b_R &=&  \zeta_Z - { c_w\over s_w} {(s - m_Z^2 )\over s}
\zeta_{AZ}  \ .
\eeqa{abRforms}
The expressions for $a_L$ and $a_R$ include the kinematic factors
\beq
 k_Z \delta m_Z + k_h \delta m_h = \half  \delta \biggl[   {1\over
  ( s-m_Z^2)^2}\cdot  {k\over \sqrt{s} } \cdot (2 + {E_Z^2\over m_Z^2}
 )\biggr] \ .
\eeqn
The expansions of these expressions in terms of the $c_I$ are given in
Appendix A.   Note, in particular, that, up to parameters that have 
already been  constrained as explained in the previous sections,   $\eta_Z =
-\half c_H$ and $\zeta_Z = c_w^2 (8 c_{WW})$.  Then, in principle, the 
 the percent-level
constrants on the $a$ and $b$ coefficients will become percent-level
constraints of on the parameters $c_H$ and $c_{WW}$.   At this point,
we have put constraints on all of the EFT parameters that contribute
to the cross section for $\ee\to Zhh$ except for the parameter $c_6$
that we hope to determine from this reaction.

\begin{table}
\begin{center}
\begin{tabular} {lcccccc}
  &   250 GeV \\ \hline
$c_I$      &  prec. EW & + $WW$ & + LHC & + $Zh$ & ILC 250 &  
\\ \hline
$c_T$     &  0.011     &   0.051    &  0.051  &  0.048  &   0.052         &                \\
$c_{HE} $ &  0.043    &    0.026    &  0.085  &  0.047  &    0.055      &                \\
$c_{HL} $ &  0.042    &    0.035    &  0.035  &  0.032  &    0.039      &                \\
$c^\prime_{HL}$ &  $-$ & 0.028    &  0.028  &  0.028 &     0.047    &                \\
$8c_{WB} $ &   $-$     &    0.078 &  0.080  &  0.076  &     0.090     &                \\
$8c_{BB}  $ &  $-$      &  $-$        &   0.20  &  0.16 &      0.11    &                \\
$8c_{WW} $ &  $-$     &    $-$       &   0.21  &  0.13 &     0.13      &                \\
$c_H$ &       $-$      &   $-$       &   $-$    &   1.12  &   1.20
&     \\ \\
  &   500 GeV \\ \hline
$c_I$    &  prec. EW & + $WW$ & + LHC & + $Zh$ & ILC 500 &  250+500 \\ \hline
$c_T$     &  0.011     &   0.046   &  0.047  &  0.041  &   0.037     & 0.030  \\
$c_{HE} $ &  0.043    &    0.015    &  0.077  &  0.040  &    0.010        &  0.009 \\
$c_{HL} $ &  0.042    &    0.030   &  0.030  &  0.027  &    0.016      &    0.013            \\
$c^\prime_{HL}$ &  $-$ & 0.027    &  0.028  &  0.026 &     0.014    &   0.011             \\
$8c_{WB} $ &   $-$     &    0.070 &  0.072  &  0.067 &     0.052     &     0.041          \\
$8c_{BB}  $ &  $-$      &  $-$        &   0.20  & 0.15&      0.088    &      0.062         \\
$8c_{WW} $ &  $-$     &    $-$       &   0.21  &  0.11 &     0.044   &      0.039          \\
$c_H$ &       $-$      &   $-$       &   $-$    &   4.78   &  1.24 &    0.65 \\
\end{tabular}
\end{center}
\caption{1 $\sigma$ uncertainties, in \%, on EFT coefficients at
              different stages of this analysis. As more information
              is included, more parameters can be added to the fit.
              Parameters that are not yet included are set to 0 and
              marked in the table with $-$.  1st column:
              precision electroweak only (7 parameter  fit);  2nd
              column: add  $\ee\to WW$ (10 parameter fit); 3rd column:
             add LHC measurements (12 parameter fit); 4th column: add
             $\ee\to Zh$ cross section, angular distribution, and
             polarization asymmetry (13 parameter fit); 5th column: 
              add $\ee\to \nu\bar \nu h$ and all $\sigma\cdot
              BR$ measurements, as described in Section 7 and in
              \cite{ImpHiggs} (22 parameter fit).    In the top half of the table, the
              $\ee$ data is the expectation for 2000~fb$^{-1}$ at
              250~GeV.
              In the bottom half of the table,  the
              $\ee$ data is the expectation for 4000~fb$^{-1}$ at
              500~GeV. The last column in the bottom half shows the
              result from the full ILC program at 250 and 500~GeV. }
 \label{tab:EFTconstraints}
\end{table}

  Table~\ref{tab:EFTconstraints} shows the 1~$\sigma$ errors on the EFT
 parameters obtained from the various stages of our fit.  The first
 four columns of the table show the results from the fits described up
 to this point. The fits have increasing numbers of parameters, from 7
 parameters in the precision electroweak fit to 22 parameters in the
 full ILC fit.   In
 each fit, we set the parameters not yet included to zero.  The analysis
 of this paper concentrates on 500~GeV measurements, but we also show
for reference 
 the fit results for 250~GeV measurements.  The Table shows the
 progression that we have explained in this paper:  Precision
 electroweak fixes three EFT coefficients, taken here to be $c_T$,
 $c_{HE}$, and $c_{HL}$, to below the $10^{-3}$ level.   Measurement
 of $\ee\to W^+W^-$ adds constraints on $c^\prime_{HL}$ and $8c_{WB}$.
 The LHC measurements of ratios of branching ratios constrain two additional
 linear combinations of the dimension-6 terms involving squares of
 field strengths and thus provide significant constraints on 
 $8c_{BB}$ and $8c_{WW}$.   Finally, adding information from $\ee\to
 Zh$ sharpens all of these constraints while also constraining the
 coefficient $c_H$.

These constraints, however, are not yet sufficiently powerful to
achieve our goal in this paper.  The problem comes from the fact that, although
the errors on $c_{HE}$,  $c_{HL}$, and $c^\prime_{HL}$ are quite
small, these parameters appear in  \leqn{abLforms} and \leqn{abRforms}
with very large coefficients, of order $2s/m_Z^2\sim 60$. This
limits the power of these equations to constrain $c_H$.  The
uncertainty on $c_H$ resulting from the analysis described so far
is about 5\%, as shown in the first entry in the last line of Table~\ref{tab:EFTconstraints}.   This
 is already consistent with
our approximation of ignoring terms of other $c_H^2$ and other terms
quadratic in  the EFT coefficients.  However, this constraint is weaker
than what we need for the determination of the triple Higgs coupling.
The constraint on
 $c_{WW}$, which comes from
the angular distribution  and polarization asymmetry in $\ee\to Zh$,
is already quite strong.

It should be noted that a similar analysis at 250~GeV, where the coefficients of the
contact terms are smaller by a factor of 4, gives a much stronger
constraint on $c_H$.   This is shown in the first entry in the last
line  of the top half
of   Table~\ref{tab:EFTconstraints}.    One can see from the sixth
column in the bottom half of Table~\ref{tab:EFTconstraints} that a
combined analysis of 250~GeV and 500~GeV data is especially powerful
to  constrain the effects of the contact interactions. 

In any event, it is also possible to improve the constraint on the
parameter $c_H$ by including additional information from $\ee$ Higgs reactions.
   In the
next Sections~6 and 7, we will explain how to improve our fit using
information from the $W$ fusion reaction $\ee\to \nu\bar\nu h$ and
from the Higgs decay partial widths.   After we add this information,
the fit results will evolve further to those shown in the fifth and
sixth columns of Table~\ref{tab:EFTconstraints}. 

The analysis  \cite{Ogawa} also considers the addition of a third,
CP-violating,
term in the effective Lagrangian,
\beq
\Delta \L = \half {\tilde b\over v}\, h\,
Z_{\mu\nu} {\tilde Z}^{\mu\nu} \ . 
\eeqn
It is found that the same data set constrains the coefficient $\tilde
b$ to be less than 1\%.    This is the final piece of information that
we need to demonstrate that---if significant CP-violating terms are not actually
generated by new physics---the possibility of CP-violating operators
does not affect the uncertainties estimated in our analysis.

\section{Constraints from $\ee\to \nu\bar\nu h$}

To obtain additional constraints on $c_H$, we now turn to the process
$\ee\to \nu\bar\nu h$. 
 Unlike $\ee\to Zh$, it is not possible to measure this
total cross section directly. But still, this process plays an important role in the
extraction of Higgs boson partial widths from $\ee$ data.

\begin{figure}
\begin{center}
\includegraphics[width=0.80\hsize]{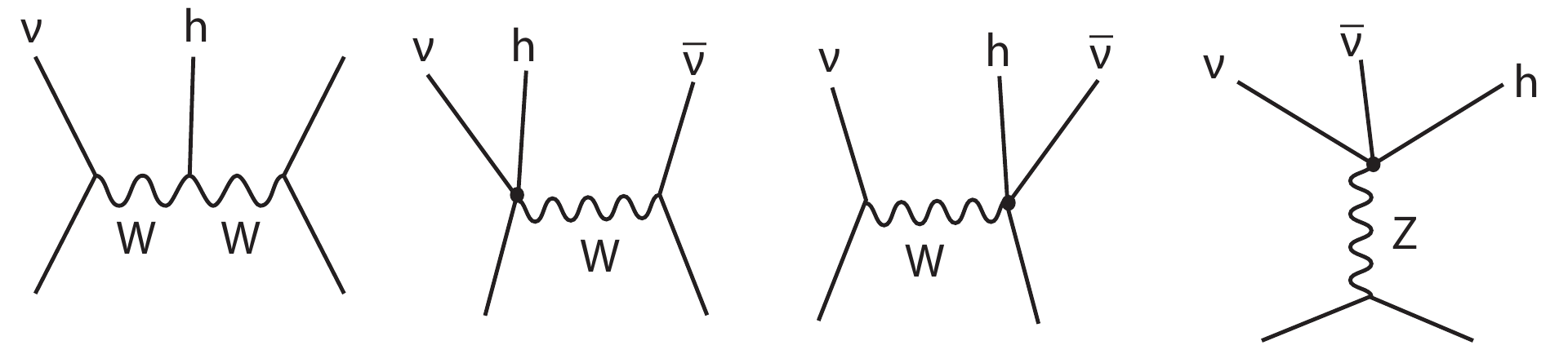}
\end{center}
\caption{Feynman diagrams contributing to the amplitudes for $\ee\to
  \nu\bar\nu h$.}
\label{fig:nunuh}
\end{figure}

The Feynman diagrams for $\ee\to\nu\bar\nu h$ are shown in 
Fig.~\ref{fig:nunuh}.  There is one helicity amplitude, for 
$e^-_Le^+_R \to \nu_L \bar \nu_R h$.  The first diagram is the one that appears at
tree level in the SM.   The additional three diagrams involve contact
interactions proportional to $c_{HL}$ and $c^\prime_{HL}$.    There
are further contributions  from the process $\ee\to Zh$, $Z \to
\nu\bar\nu$, but these are important only when the final $Z$ is close
to its mass shell.  We will ignore them, and, more generally, we will ignore
interference between the $W$ fusion reaction and the $\ee\to Zh$
reaction.

The first diagram shown in Fig.~\ref{fig:nunuh} has the value
\beqa
i\M &=& i {g_W^2\over 2} \biggl\{  { 2 m_W^2\over v}(1 + \eta_W)
g^{\mu\nu} -  {2\over v} \zeta_W (q_1\cdot q_2 g^{\mu\nu} - q_1^\nu
q_2^\mu) \biggr\}  \CR
 & & \hskip 0.5in    \cdot  {1\over (q_1^2 - m_W^2)(q_2^2 - m_W^2)} \cdot
\bar u_L(\nu) \gamma_\mu u_L(e^-) \ \bar
 v_R(e^+) \gamma_\nu v_R(\bar\nu) \, 
\eeqan
where $q_1$, $q_2$ are the momenta of the two off-shell $W$ bosons.
Including also the various contact interactions, the full expression
for this amplitude is
\beqa
i\M &=& i {g_W^2\over 2} \biggl\{  { 2 m_W^2\over v}(1 + \eta_W)
g^{\mu\nu} -  {2\over v} \zeta_W (q_1\cdot q_2 g^{\mu\nu} - q_1^\nu
q_2^\mu) \CR
& & \hskip 0.6in    + 2 c^\prime_{HL} \bigl( {q_1^2 - m_W^2 + q_2^2 -
m_W^2\over 2m_W^2}\bigr) \biggr\} \CR
 & & \hskip 0.3in    \cdot  {1\over (q_1^2 - m_W^2)(q_2^2 - m_W^2)}  \cdot
 \bar u_L(\nu) \gamma_\mu u_L(e^-) \ \bar v_R(e^+) \gamma_\nu
 v_R(\bar\nu)\CR
 & & \hskip 0.1in -{ g^2\over c_w^2} {g^{\mu\nu}\over v (s - m_Z^2)} (c_{HL} -
 c_{HL}^\prime) \cdot \bar u_L(\nu) \gamma_\mu v_R(\bar \nu) \ \bar v_R(e^+) \gamma_\nu
 v_R(\bar\nu) \ . 
\eeqa{nnhamp}

It is not straightforward to quote analytic results for the dependence
of the total cross section on the EFT parameters.  However, we can
integrate the expression \leqn{nnhamp}
 over the 3-body phase space numerically to compute the fully
 polarized cross section.  We obtain, for 
$\sqrt{s} = 250$~GeV,
\beqa
\sigma/(SM) &=&  1 + 2\eta_W - 2\delta v + 2 \delta g_W -  1.6 \delta m_W - 3.7 \delta
m_h  \CR & & \hskip 0.2in  - 0.22 \zeta_W - 6.4 c^\prime_{HL}- 0.37
  (c_{HL} - c^\prime_{HL}) \ , 
\eeqa{nnhone}
for $\sqrt{s} = 350$~GeV,
\beqa
\sigma/(SM) &=&  1 + 2\eta_W - 2\delta v + 2 \delta g_W -  1.2 \delta
m_W - 2.0 \delta
m_h  \CR & & \hskip 0.2in  - 0.32 \zeta_W - 7.5 c^\prime_{HL}- 0.28
  (c_{HL} - c^\prime_{HL}) \ , 
\eeqa{nnhtwo}
for $\sqrt{s} = 380$~GeV,
\beqa
\sigma/(SM) &=&  1 + 2\eta_W - 2\delta v + 2 \delta g_W - 1.1 \delta
m_W - 1.7 \delta
m_h  \CR & & \hskip 0.2in  - 0.34 \zeta_W - 7.8 c^\prime_{HL}- 0.26
  (c_{HL} - c^\prime_{HL}) \ . 
\eeqa{nnhthreeplus}
and for $\sqrt{s} = 500$~GeV,
\beqa
\sigma/(SM) &=&  1 + 2\eta_W - 2\delta v + 2 \delta g_W -  0.85 \delta
m_W - 1.2 \delta
m_h  \CR & & \hskip 0.2in  - 0.39 \zeta_W - 8.8 c^\prime_{HL}- 0.19
  (c_{HL} - c^\prime_{HL}) \ . 
\eeqa{nnhthree}
Each expression contains $(-c_H - a (8 c_{WW})\,) $, with the first term
coming from $\eta_W$ and the second from $\zeta_W$.   The coefficient
$a$ of $\zeta_W$ increases slowly with center of mass energy.   Thus,
measurements $\sigma \cdot BR$ for $WW$ fusion to a Higgs boson and
then to a given final state can constrain the parameters $c_H$ and
$c_{WW}$ in the context of a global fit to Higgs boson data.

In the second line of each of these expressions, the second term comes
from the diagrams with  contact interactions and $t$-channel $W$
exchange.
The numerical coefficients in these $c^\prime_{HL}$ terms are large and increase with
center of mass energy, just as we saw for the contact contributions 
in $\ee\to Zh$.   However, now
there is an interesting possibility.   If the cross sections for both
processes are measured, the contact interaction coefficients are
overdetermined and can be constrained even more strongly than they are
from precision electroweak data.   We will see in Section 8 that this
is indeed the case.

Since the total cross section for $\ee\to \nu\bar\nu h$ cannot be
measured directly, we must consider the formulae \leqn{nnhone},
\leqn{nnhtwo}, and \leqn{nnhthree} in conjuction with formulae for
Higgs decay processes.  We develop these formulae in the next section.

\section{EFT formalism for general Higgs boson couplings}

In the process of answering the main issue of this paper, we have
already come
very close to assembling the complete set of formulae that we need to
represent  general Higgs boson cross sections at $\ee$ colliders in
terms of EFT coefficients.   In this
section, we derive the remaining formulae needed for such an
analysis.  These are the formula for the various Higgs decay widths. 
 The implications of the formalism of this paper for the
extraction of Higgs couplings at $\ee$ collider will be discussed in
a companion paper~\cite{ImpHiggs}. 

In Section 4, we derived expansions for two of the minor decay
amplitudes, $h\to \gamma\gamma$ and $h\to Z\gamma$.  What remains is
to derive formulae for the major Higgs boson decay
amplitudes to fermions, $WW^*$, and $ZZ^*$.

\subsection{Higgs decay to fermions and gluons}

At the level of this tree-level analysis, the appropriate treatment of
Higgs decays to fermions is very simple.  For definiteness, consider
the case of $h\to \tau^+\tau^-$.   Deviations in the Higgs
couplings from the SM expectation are generated by  the dimension-6 operator
\beq
\Delta \L =   -
c_{\tau\Phi} { y_\tau\over
  v^2} (\Phi^\dagger \Phi) \bar L_3 \cdot \Phi \tau_R + h.c. \ ,
\eeq{taushift}
where 
$y_\tau$ is the bare Yukawa coupling.  Then 
\beq
  m_\tau =   {y_\tau v \over \sqrt{2}} (1 + \half  c_{\tau\Phi})
\eeqn
Substituting $m_\tau$ for $y_\tau$ using this formula and including
the Higgs field strength renormalization from \leqn{Zvals}, the $\tau$
couplings to the Higgs boson becomes
\beq
\Delta \L =   -  m_\tau  \bar\tau \tau \cdot (1 - \half c_H + 
c_{\tau\Phi}) \cdot {h\over v}  \ .  
\eeqn
The variation of the Higgs width is then
\beq
\delta \Gamma(h\to \tau^+ \tau^-)  =   1 - c_H + 2 c_{\tau\Phi} +
\delta \ , 
\eeqn
where $\delta = 2\delta m_\tau + \delta m_h - 2 \delta v$.  For
simplicity, we will absorb this term into $c_{\tau\Phi}$.

A similar logic applies to the Higgs boson couplings to $b$, $c$,
$\mu$, and other fermions.  Then, we will write
\beqa
\delta \Gamma(h\to b \bar b) & = &  1 - c_H + 2 c_{b\Phi}\CR
\delta \Gamma(h\to c \bar c) & = &  1 - c_H + 2 c_{c\Phi}\CR
\delta \Gamma(h\to \tau^+ \tau^-) & = &  1 - c_H + 2 c_{\tau\Phi} \CR
\delta \Gamma(h\to \mu^+\mu^-) & = &  1 - c_H + 2 c_{\mu\Phi}  \ .
\eeqan
QCD corrections provide factors that commute with the effect of
dimension-6 operators and so do not affect these formulae.   Mixed
QCD-electroweak corrections will give loop-level corrections to
these formulae.

The effect of dimension-6 operators on the partial width for $h\to gg$
is more complex.   The first contribution to this width in the SM
comes at the loop level.   Dimension-6 operators correct this
expression through a tree-level contribution proportional to the
coefficient $c_{GG}$ of a gluonic operator similar to that for
$c_{WW}$, and through corrections to the SM loop diagrams, for example, from
$c_{t\Phi}$.   Fortunately, for an on-shell Higgs boson, 
 it is a  good approximation to summarize
all of these effects as an effective coupling of the form
\beq
\delta \L =   {\cal A}  {h\over v}  G_{\mu\nu} G^{\mu\nu} \ . 
\eeqn
In fitting Higgs couplings, we will write 
\beq
\delta \Gamma(h\to gg)  =   1 - c_H + 2 c_{g\Phi}\ , 
\eeqn
letting the parameter $c_{g\phi}$ stand in for all of the effects just
described.

A full description of the $h\to gg$ width in the EFT formalism would include the dependence
of this partial width on the canonical EFT parameters $c_{GG}$,
$c_{t\Phi}$, and $c_{tG}$~\cite{Polish}, with small corrections from
other dimension-6 operators.  That discussion is beyond the scope of
this paper.  The leading effects can be disentangled
by measurements of Higgs emission from $t\bar t$, Higgs production in
$pp$ collisions at high $p_T$, and top quark pair production at high
energy.  A part of this analysis is given in \cite{HiggsatlargepT,Ilnicka}.

\subsection{Higgs decay to  $WW^*$ and $ZZ^*$}

The Higgs decay widths to $WW^*$ and $ZZ^*$ also bring in new EFT
vertices.  However, in this case, the new terms can be constrained by additional
precision electroweak measurements.

\begin{figure}
\begin{center}
\includegraphics[width=0.70\hsize]{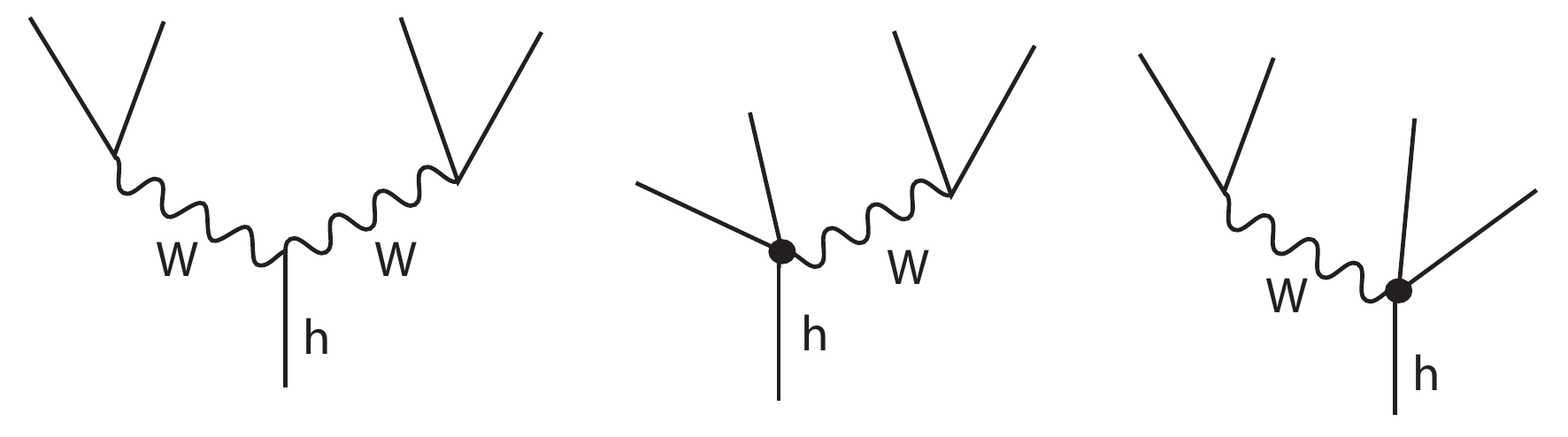}
\end{center}
\caption{Feynman diagrams contributing to the amplitudes for $h\to WW^*$.}
\label{fig:hWWdiagrams}
\end{figure}

As a first step into this analysis, consider  a model of  $h\to WW^*$ in
which the $W^-$ converts only to $e^-\bar\nu$ and the $W^+$ converts
to only to 
$e^+\nu$.  In this case, the $W$ width would be 
\beq
\Gamma_{W,simple} = { g_W^2 m_W\over 48\pi}  =  { g^2m_W\over 48\pi} (1
+ 2\delta g + \delta m_W + 2 c'_{HL} +  \delta Z_W) \ . 
\eeq{simpleWwidth}
For an off-shell $W$, we will use the propagator
\beq
 D(q) = {1/(q^2 - m_W^2
+ i q^2 (\Gamma_W/m_W) ) }  
\eeqn
with a $q^2$-dependent width. 

It is straightforward to compute the rate of the $h\to WW^*$ decay in
this model. 
The Feynman diagrams are shown in
Fig.~\ref{fig:hWWdiagrams}.   Note that, in addition to the usual SM
diagram, there are contributions from the contact interaction
proportional to $c^\prime_{HL}$.   The decay amplitude is 
\beqa
 i\M &=& i {g_W^2\over 2}\biggl\{  {2m_W^2\over v} g^{\mu\nu} \biggl[
 (1+\eta_W)D(q_1^2) D(q_2^2)  +{ c^\prime_{HL} \over 2m_W^2}\bigl(D(q_1^2)
  + D(q_2^2) \bigr)\biggr] \CR
& & \hskip 0.1in - {2\over v}\zeta_W (q_1^2\cdot q_2^2 g^{\mu\nu} -
q_1^\nu q_2^\mu) \biggr\} \CR
& & \hskip 0.4in  \bar u_L(e^-)\gamma_\mu v_R(\bar\nu) \ \bar
u_L(\nu)\gamma_\nu v_R(e^+)  \ , 
\eeqan
where $q_1$, $q_2$ are the momenta of the $W^-$ and $W^+$.
Integrating this expression over phase space and using
\leqn{simpleWwidth} to simplify the numerator, we find
\beqa
\Gamma/(SM) &=&  1 + 2\eta_W - 2\delta v -  11.7 \delta
m_W+ 13.6\delta
m_h  \CR & & \hskip 0.2in  - 0.75 \zeta_W - 0.88 C_W + 1.06
\delta \Gamma_W \ ,
\eeqa{Wdep}
where we have written  $C_W = c^\prime_{HL}$. 
There is a partial  cancellation between the factors of
$c^\prime_{HL}$ that appear explicitly due to the contact interactions
and the factors that appear in $\Gamma_W$ through
\leqn{simpleWwidth}. 

In reality, the $W$ boson can decay to all of the SM $SU(2)$ doublets
except $(t,b)$.  This brings in additional $c^\prime_{HX}$
coefficients for the first and second quark generations.
Fortunately, these new coefficients appear only in the same
combination that appears in the full $W$ width.  Let
\beq
     C_W =  \sum_X  c^\prime_{X} {\cal N}_X \ / \  \sum_X {\cal N}_X 
   \ ,
\eeqn
where $X$ runs over the five SM doublets that appear in $W$ decays,
$c^\prime_{X}$ is the coefficient of the operator similar to that
multiplying $c^\prime_{HL}$, and ${\cal N}_X$ is the number of color
states for that doublet, including the QCD radiative correction.
Then, including all first-order EFT corrections,  the $W$ width is given by 
\beq
\Gamma_{W} =  { g^2m_W\over 48\pi} (\sum_X {\cal N}_X) \cdot (1 + 2 \delta
g + \delta m_W + 
\delta Z_W
+ 2 C_W) \ . 
\eeq{Wwidth}
The expression \leqn{Wdep}
remains valid, but with $c^\prime_{HL}$ replaced by
$C_W$.   We can constrain the value of $C_W$ by a measurement of the
$W$ total width, and then \leqn{Wdep} becomes an additional constraint
on the EFT parameters $c_H$ and $c_{WW}$.

It is also striking that the expression \leqn{Wdep} shows a very
strong dependence on the masses of the $W$ boson and the Higgs boson.
The improvements in these quantities expected from LHC and ILC and listed in
Table~\ref{tab:PEW} will be important to make use of the Higgs boson
width to $WW^*$ in a global fit to the Higgs boson   couplings.

The analysis of $h\to ZZ^*$ is formally quite similar, but there is
some additional
bookkeeping to do.   Write the SM  coupling of one chiral
flavor $X$ to the $Z$ boson as 
\beq
\Delta \L =   {g\over c_w}     Q_{ZX} \  Z_\mu  \bar X \gamma^\mu X \  ,
\eeqn
where $Q_{ZX} = I^3_X - s_w^2 Q_X$, with $I^3_X$, $Q_X$ the weak isospin and the electric charge of $X$.
The contact interactions yield an additional direct coupling 
\beq
\Delta \L =   {g\over c_w}     c_{X}  \  Z_\mu  \bar X \gamma^\mu X \ 
\bigl(1 + 2 {h\over v} + \cdots \bigr) \ ,
\eeq{Zcontact}
introducing a new parameter $c_X$ for each chiral flavor.  When we
include this effect and all other  first-order EFT corrections, the
coupling  of the $Z$ to $X\bar X$ is modified
to
\beqa
  g_X &=& {g\over c_w}\bigl[ Q_{ZX} (1 + c_w^2 \delta g + s_w^2 \delta
  g' + \half \delta Z_Z) \CR & & \hskip 0.2in + Q_X ( 2 s_w^2 c_w^2 (\delta g -\delta g') + s_w
  c_w \delta Z_{AZ} ) + c_{X} \bigr] \ .
\eeqan
Then the total $Z$ width becomes 
\beqa
 \Gamma_Z &=&  {g^2 m_Z\over 24\pi c_w^2}  (\sum_X  Q_{ZX}^2
 {\cal  N}_X ) \cdot \biggl[ (1 + 2 c_w^2 \delta g + 2 s_w^2 \delta g'
 +\delta m_Z+ \delta Z_Z )\CR
& & \hskip 0.1in  + {\sum_X Q_{ZX} Q_X{\cal N}_X\over \sum_X Q_{ZX}^2 {\cal N}_X }
  ( 4 s_w^2 c_w^2 (\delta g - \delta g') + s_w c_w \delta Z_{AZ} )
\biggr] \cdot (1 + 2 C_Z) \ , 
\eeqan
where
\beq
     C_Z = { \sum_X \  c_X Q_{ZX} {\cal N}_X \over \sum_X \ Q_{ZX}^2 {\cal
       N}_X }\ . 
\eeqn
For the $Z$ decaying to SM fermions,
\beq
  \sum_X Q_{ZX}^2 {\cal N}_X = 3.75 \ , \quad  \sum_X Q_{ZX} Q_X {\cal
    N}_X  = 1.99  \ .
\eeqn

The contact interaction also affects the $h\to ZZ^*$ decay by adding
additional contact diagrams similar to those in
Fig.~\ref{fig:hWWdiagrams}.  We find
\beqa
\Gamma/(SM) &=&  1 + 2\eta_Z - 2\delta v -  13.8\delta
m_W + 15.6\delta
m_h  \CR & & \hskip 0.2in  - 0.50 \zeta_Z - 1.02 C_Z + 1.18 
\delta \Gamma_Z \ . 
\eeqa{Zdep}
So, here again, there is an extra EFT parameter, but it  can be
controlled by measurement of the $Z$ total width.

The conclusions of this section are summarized in Appendix A.

\section{The total cross section for $\ee\to Zhh$} 

\begin{figure}
\begin{center}
\includegraphics[width=0.80\hsize]{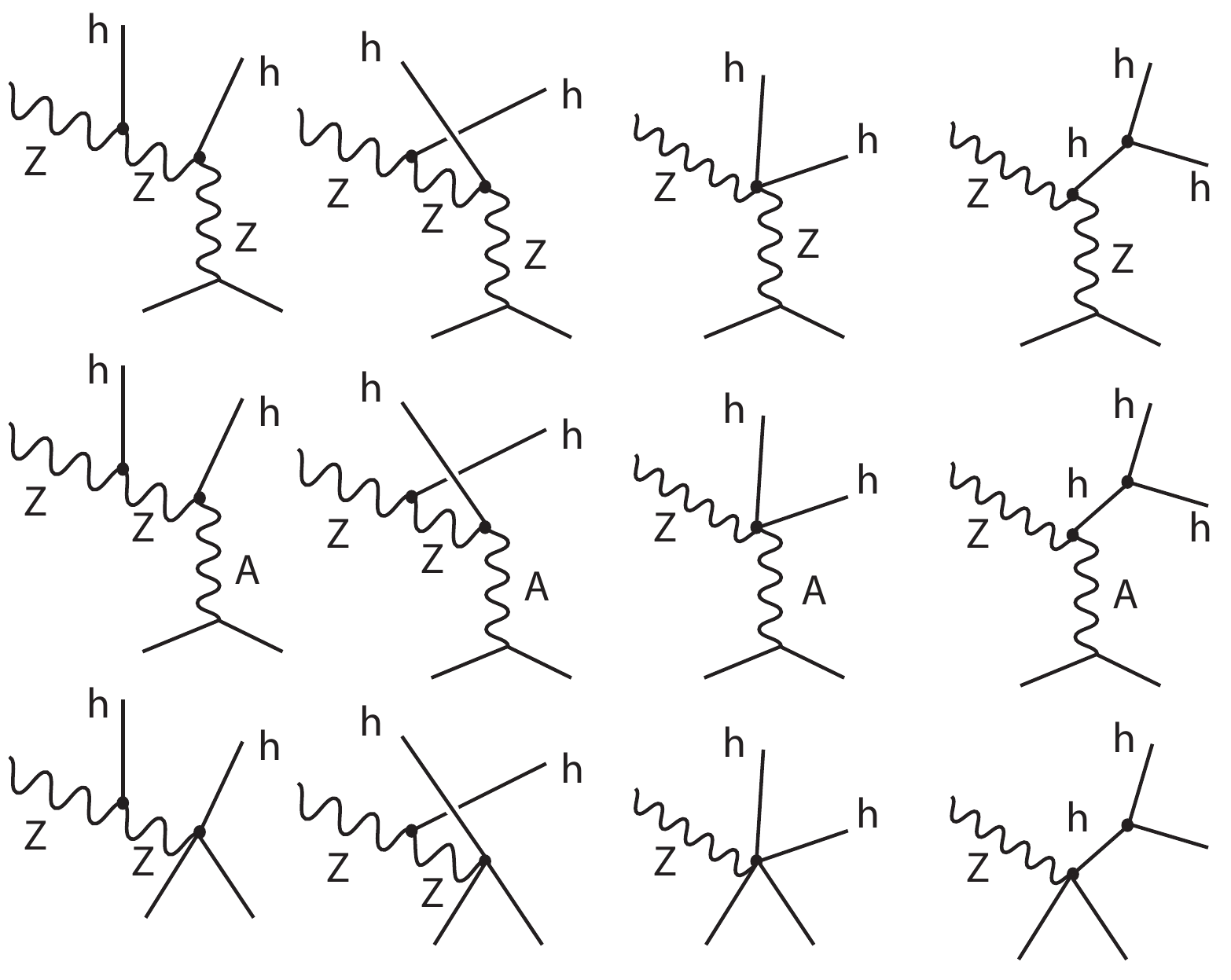}
\end{center}
\caption{Feynman diagrams contributing to the amplitudes for $\ee\to Zhh$.}
\label{fig:Zhhdiagrams}
\end{figure}

We are now ready to describe the derivation of the parameter $c_6$
from the value of the total cross section for $\ee\to Zhh$. 
  The tree-level Feynman diagrams for this process  are shown in
Fig.~\ref{fig:Zhhdiagrams}.    Evaluating these diagrams and
numerically
integrating over three-body phase space, we will obtain expression of
a form similar to our cross section formulae for $\ee\to \nu\bar\nu
h$.

The diagrams in the first row of
Fig.~\ref{fig:Zhhdiagrams} are those of the Standard Model.  However,
in our EFT formalism, all $ZZh$ vertices also include renoramalization
of all Higgs vertices by $\delta Z_h$ and new structures proportional
to $\zeta_Z$.  The last diagram in this row contains the modification
of the triple Higgs coupling proportion to  $c_6$ but also the
additional vertex structure  from the term proportional to $\theta_h$
in \leqn{Higgsform}.   The diagram in the second row make use of the
$\zeta_{AZ}$ term that converts $A$ to $Z$ while emitting one or more
Higgs bosons.  Recall that kinetic mixing between $A$ and $Z$ in the
$s$-channel propagator has already been taken into account in the
parameters $g_L$, $g_R$.   Diagrams with kinetic mixing beyond the
first vertex are of order $c^2_I$ and can be ignored.    The diagrams
in the third row involve the contact interactions proportional to
$(c_{HL}-c^\prime_{HL})$ and $c_{HE}$.   In all, there are many
opportunities for EFT coefficients other than $c_6$ to influence the
value of this cross section.

The amplitude for $\ee\to Zhh$ depends on the initial beam
polarization and on the final polarization state of the $Z$.  We
compute the cross section at $\sqrt{s} = 500$~GeV for definite choices of the initial beam
polarization and summed over $Z$ helicities.   For a fully polarized
initial state $e^-_Le^+_R$, we find
\beqa
\sigma/(SM) &=& 1 + 2 \delta g_L +  1.40 \eta_Z + 1.02 \eta_{ZZ} +
18.6 \zeta_Z + 24.8 \zeta_{AZ} \CR
  && \hskip 0.2in + 0.56 \eta_h - 1.58 \theta_h +  108.3 (c_{HL} +
  c^\prime_{HL}) \CR
  & & \hskip 0.2in  - 3.9 \delta m_h + 3.5\delta m_Z  \ .
\eeqa{LZhh}
For a fully polarized initial state $e^-_Re^+_L$, we find
\beqa
\sigma/(SM) &=& 1 + 2 \delta g_R +  1.40 \eta_Z + 1.02 \eta_{ZZ} +
18.6 \zeta_Z - 28.7 \zeta_{AZ} \CR
  && \hskip 0.2in + 0.56 \eta_h - 1.58 \theta_h - 125.5 c_{HE}\CR
  & & \hskip 0.2in  - 3.9 \delta m_h + 3.5 \delta m_Z  \ .
\eeqa{RZhh}
For an unpolarized $\ee$ initial state, we find
\beqa
\sigma/(SM) &=& 1 + 1.15\delta g_L + 0.85 \delta g_R +   1.40 \eta_Z + 1.02 \eta_{ZZ} +
18.6 \zeta_Z + 2.0 \zeta_{AZ} \CR
  && \hskip 0.2in + 0.56 \eta_h - 1.58 \theta_h +  62.1 (c_{HL} +
  c^\prime_{HL}) - 53.5 c_{HE}\CR
  & & \hskip 0.2in  - 3.9 \delta m_h + 3.5 \delta m_Z  \ .
\eeqa{UZhh}
These equations are rewritten with some convenient rearrangement of terms in
Appendix A. 

We find the dependence on EFT parameters shown in this equation to be
quite surprising.  It is well known that the dependence of the $\ee\to
Zhh$ cross section on the triple Higgs coupling is weak.    Here, that
dependence appears in the coefficient of $\eta_h =  c_6 + \cdots$.
The relation 
\beq 
    \sigma/(SM) = 1 +  0.56 c_6 + \cdots 
\eeq{csixdep}
agrees with \cite{ILChh} and earlier studies.   What is remarkable is
that the dependence on other parameters is much larger.  We might pay
particular attention to the dependence on $c_H$ and $c_{WW}$, the two
parameters that are only fixed by single-Higgs production processes.
The parameter  $c_H$ appears in $\eta_Z$, $\eta_{ZZ}$, $\eta_h$, and
$\theta_h$.  The parameter $c_{WW}$ appears in $\zeta_Z$ and
$\zeta_{AZ}$; we omit a further dependence from the independently
constrained
$\delta g_{L.R}$.  The
sum of these terms gives (in the unpolarized case)
\beq 
 \sigma/(SM) =  1 - 4.15 c_H  + 15.1 (8c_{WW}) + \cdots 
\eeqn
The coefficients here are an order of magnitude larger than that in
\leqn{csixdep}. In addition, the parameters  $(c_{HL}+c^\prime_{HL})$
and $c_{HE}$, which are constrained by precision electroweak
measurements,
have very large coefficients, reflecting an $s/m_Z^2$ enhancement of
their contributions.   We have seen this effect already in both of the
single-Higgs boson reactions considered earlier in this paper.
  It is clear that, without precise constraints on
the EFT  parameters from all of the sources that we have discussed in
this paper, it is not possible
to convincingly attribute a measured increase in the double Higgs production cross
section to a shift in the triple Higgs coupling.

We can discuss this quantitatively using a fit to the EFT parameters
aside from $c_6$ using the inputs in Table~\ref{tab:PEW}, for
precision electroweak, and the inputs listed in Appendix B for
$W^+W^-$, and the measurement of the $a_{L,R}$ and $b_{L,R}$ parameters
in $\ee\to Zh$.     This fit involves 13 parameters, the 4 SM
parameters and the 9 EFT coefficients introduced in Section 2.   The
fit results for the relevant $c_I$ parameters have already been shown
in Table~\ref{tab:EFTconstraints}. 
This fit leads to the following
values for the root-mean-square errors (in \%) on EFT coefficients:
\beq
\begin{tabular}{lc|lc}
$A$   &    $ [ < A^2> ]^{1/2}$  &   $A$    &  $ [ < A^2> ]^{1/2}$  \\ \hline
$c_H$ &              4.8           &   $(c_{HL}+c^\prime_{HL})$
&      0.048\\
$(8c_{WW})$ &      0.11   & $c_{HE}$ &  0.040 \\
$(-4.15 c_H + 15.1 (8 c_{WW}) )$ &   21  &     
 $62.1 (c_{HL} + c^\prime_{HL}) - 53.5 c_{HE} $ &       4.9
\end{tabular}
\eeq{finaltable}
We find for the root-mean-square uncertainty in the complete
right-hand side of \leqn{UZhh}, omitting the dependence on $c_6$
\beq
    [\VEV{(\delta \sigma)^2}]^{1/2} =  14\% \ .
\eeqn
This means that a measurement of $c_6$ from the cross section for
$\ee\to Zhh$ will be subject to a  28\%  systematic uncertainty from
the uncertainties in the other EFT parameters.  So the logic that we
have described is in principle valid, but it leads to a very large
uncertainty from other new physics effects in the determination of $c_6$.

 We pointed out at the end of Section~5 that this problem can be solved
by adding data
the various
$\sigma\cdot BR$ measurements possible with $\ee\to\nu\bar\nu h$,
together with information from $\sigma\cdot BR$ measurements in
$\ee\to Zh$.   Given the absolute measurement of the total cross section for
$\ee\to Zh$, these additional measurements fix the various new parameters
that appear in the Higgs boson decay amplitudes.   Using the fit to
these parameters, we can bootstrap the measurement of the total
cross section for $\ee\to Zh$ into a determination of the total cross
section for $\ee\to \nu\bar\nu h$ and the absolute normalization of
the partial widths $\Gamma(h\to WW^*)$ and $\Gamma(h\to ZZ^*)$.
This gives an independent way to determine $c_H$.  This method is
applied in the fits presented in the fifth and sixth columns of 
 Table~\ref{tab:EFTconstraints}, and one can see from that Table that
 it is effective.  The full set of
inputs to these
fits, and the results for Higgs boson
couplings and decay amplitudes, are described in detail in
 \cite{ImpHiggs}.   

Using  the fit described in the final column of
Table~\ref{tab:EFTconstraints}, 
including all cross sections and branching fractions that
will be measured at the ILC at 250 and 500~GeV,  the errors reported in
\leqn{finaltable}
improve to 
\beq
\begin{tabular}{lc|lc}
$A$   &    $ [ < A^2> ]^{1/2}$  &   $A$    &  $ [ < A^2> ]^{1/2}$  \\ \hline
$c_H$ &             0.65                     &
$(c_{HL}+c^\prime_{HL})$ &  0.014 \\
$(8c_{WW})$ &     0.039          & $c_{HE}$ &  0.009 \\
$(-4.15 c_H + 15.1 (8 c_{WW}) )$ & 2.8  &     
 $62.1 (c_{HL} + c^\prime_{HL}) - 53.5 c_{HE} $ &   0.85
\end{tabular}
\eeq{reallyfinaltable}
and the uncertainty in $\delta \sigma$ becomes
\beq
    [\VEV{(\delta \sigma)^2}]^{1/2} =  2.4 \%  \ .
\eeqn
At this point, the effects of other EFT coefficients contribute only
a 5\% systematic error to the determination of the parameter $c_6$,
and so this parameter can be determined from the measurement of the
$\ee\to Zhh$ cross section with high precision in a model-independent
way.

As an aside, we note that the full fits to Higgs observables give
quite an impressive improvement in the uncertainties in the
parameters $c_{HE}$, $c_{HL}$, and $c^\prime_{HL}$  from the original
precision electroweak determination.
In precision electroweak observables, the $c_{HL}$ and related
parameters 
alter the $W$ and $Z$
couplings with coefficients that are of order 1.   In the EFT
formalism, these same parameters appear as contact interactions in the
Higgs reactions, with coefficients that are enhanced by factors of
order $s/m_Z^2$.   Then the sensitivity to these factors is much
stronger.  The EFT formalism implies that the measurement of Higgs
reactions can provide more powerful tests of deviations of the
predictions of precision electroweak analysis than precision
electroweak measurements themselves.

\section{Conclusions} 

In this paper, we have assembled a complete formalism, valid at the
tree level and to linear order in the coefficients of dimension-6
operators, describing the possible new physics perturbations of the
Standard Model
predictions for precision electroweak observables, $\ee\to W^+W^-$,
and Higgs boson production and decay reactions. This formalism
requires a fit to  14 variables for the determination of the triple
Higgs coupling and an additional 7 variables for a general analysis of
Higgs decays  to Standard Model particles.   However, it provides a  completely
model-independent description of the effects of new physics that
arises at mass scales much larger than the mass of the Higgs boson.

It is challenging to fit this large number of parameters with high precision
and with systematic understanding of the constraints.  However, future
$\ee$ colliders will be up to this challenge.  We have shown that the
determination of the parameters can make use of all of the important
advantages of $\ee$ experimentation: beam polarization, the
visibility  of all relevant decay channels, and the ability to
measure over essentially all of phase space.    It is already
understood that these are powerful capabilities, but it is wonderful
to see in this analysis how these powerful measurements interlock to
provide  a rich and secure basis from which to explore for new
effects.

The analysis that we have described is particularly important for the
determination of the triple Higgs coupling.   This fundamental
quantity of the Standard Model is never seen in isolation.  It is
always studied as an interference effect, in combination with many
other particle vertices.   We might be able to measure a deviation
that could plausibly arise from a shift of the triple Higgs coupling,
but to understand definitely that this and not some other perturbation is
the cause, an analysis of the type described in this paper is
required.

It is difficult to imagine repeating the analysis presented here with
data from hadron colliders only.   The use of hadronic initial states
brings in many more unknown coefficients of dimension-6 operators,
while offering fewer tools to discriminate between their effects.
For the triple Higgs coupling, there is the additional complication
that the leading double Higgs production process, $gg\to hh$, is
loop-level in the Standard Model, which adds another layer of
complexity.

Thus, a future $\ee$ collider is not only sufficient but also
essential for a full understanding of the physics of the Higgs boson.

\Acknowledgements

We are grateful to many people with whom we have discussed this
analysis, including   Gauthier Durieux, Christophe Grojean, Jiayin Gu,
Howard Haber, Jenny List, Tomohisa Ogawa, Tomohiko Tanabe,  
Kechen
Wang, Liantao Wang, and Jacqueline Yan.
TB, SJ,  and MEP
were supported by the US Department of Energy under   
 contract DE--AC02--76SF00515.  TB was also supported by a KEK
 Short-Term
Invited Fellowship.   He thanks the KEK ILC group for hospitality
during this visit. 
   SJ was also supported by
the  National Research
Foundation of Korea under grants 2015R1A4A1042542
  and 2017R1D1A1B03030820.   KF was supported
by the Japan Society for the Promotion of Science (JSPS) under
Grants-in-Aid for Science Research 16H02173 and 16H02176.  JT was
supported by  JSPS under Grant-in-Aid 15H02083.

\appendix

\section{Expansions in small parameters used in our analysis}

In this Appendix, we list the expansions in SM coupling shifts and
$c_I$ operator coefficients used in the analysis of this paper.   The
notation is   $\delta A =   \Delta A/A$.  

Observables depend on the underlying parameters both directly, through
the coupling constants, and indirectly, through kinematic dependence
on the  masses $m_W$, $m_Z$, and $m_h$, which in turn depend on the
coupling constants.  In these formulae, we track both types of
dependence.  The variation of parameters contributing to
the boson masses and the physical couplings is controlled by
measurements of these masses and couplings that are included in our fit.

Expansions of boson field strength renormalizations:
\beqa
    \delta Z_W &=&  (8c_{WW}) \CR
    \delta Z_Z &=&   c_w^2 (8c_{WW})  +2 s_w^2 (8c_{WB} ) +
      {s_w^4/c_w^2} (8 c_{BB}) \CR
    \delta Z_A &=&  s_w^2 \biggl(  (8c_{WW})  - 2 (8c_{WB} ) +
      (8c_{BB}) \biggr) \CR
  \delta Z_{AZ} &=&  s_w c_w  \biggl(   (8c_{WW})  - (1 - {s_w^2\over
    c_w^2})(8c_{WB})  - {s_w^2\over c_w^2} (8c_{BB}) \biggr) \CR
    \delta Z_h &=&  - c_H  \ \ .
\eeqa{deltaone} 

Expansions of bare couplings:
\beqa
\delta \ [g^2 + g^{\prime 2}]^{1/2} &=&   c_w^2 \delta g + s_w^2 \delta
g'\CR
\delta\ ( gg'/[g^2 + g^{\prime 2}]^{1/2} ) &=&   s_w^2 \delta g + c_w^2
\delta g' \CR
\delta s_w &=& - c_w^2 ( \delta g - \delta g')\CR
\delta c_w &=& s_w^2 (\delta g - \delta g')
\eeqan

Expansions of physical couplings:
\beqa
\delta e &=&   s_w^2 \delta g + c_w^2 \delta g' + \half \delta Z_A \CR
\delta g_L &=& {1\over (1/2-s_w^2)} \biggl[  c_w^2 (\half + s_w^2)
\delta g - s_w^2 (\half + c_w^2) \delta g' + \half (c_{HL} +
c^\prime_{HL}) \CR
& & \hskip 0.1in + {1\over 4} c_w^2 (1 + 2 s_w^2)(8c_{WW}) - \half
s_w^2 (1 - 2s_w^2) (8c_{WB}) - {1\over 4} {s_w^4\over c_w^2}
(1+2c_w^2) (8
c_{BB}) \biggr] \CR
\delta g_R &=&  - c_w^2 \delta g + (1+c_w^2) \delta g' - {1\over
  2s_w^2} c_{HE} \CR
  & & \hskip 0.1in - \half c_w^2 (8c_{WW}) + c_w^2 (8c_{WB}) + \half
  {s_w^2\over c_w^2} (1+c_w^2)(8c_{BB}) \CR
\delta g_W &=& \delta g + c^\prime_{HL} + \half (8c_{WW}) \CR
\delta g_Z &=& (1 + s_w^2) \delta g - s_w^2 \delta g' + \half \delta
Z_Z + {s_w\over c_w} \delta Z_{AZ} 
\eeqan 

Expansions of boson masses:
\beqa
\delta m_W &=&   \delta g + \delta v + \half \delta Z_W\CR
\delta m_Z  &=&   c_w^2 \delta g + s_w^2 \delta g' + \delta v   -
\half c_T + \half \delta Z_Z \CR
\delta m_h  &= &  \half \delta \bar\lambda + \delta v + \half \delta
Z_h
\eeqan

Expansions of precision electroweak observables:
\beqa
\delta \alpha^{-1}  &=& - 2 \delta e \CR
\delta G_F &=&  - 2 \delta v + 2 c^\prime_{HL}\CR
\delta A_\ell &=& {4 g_L^2 g_R^2 (\delta g_L - \delta g_R)\over (g_L^2
  +g_R^2)(g_L^2 - g_R^2)} \CR
\delta \Gamma_\ell &=&  \delta m_Z +  { 2 g_L^2 \delta g_L + 2 g_R^2
  \delta g_R \over (g_L^2 + g_R^2)} \CR
\delta\Gamma_{W,\ell} &=&  \delta m_W + 2 \delta g_W\CR
\delta \Gamma_W &=& 2\delta g + \delta m_W + \delta Z_W + 2 C_W \CR
\delta \Gamma_Z &=&   2c_w^2 (1 +2 {\cal Q} s_w^2)  \delta g + 2s_w^2
(1 - 2 {\cal Q} c_w^2) \delta g' + \delta m_Z
+ \delta Z_Z \CR
 & & \hskip 0.1in + {\cal Q} 
 s_wc_w\delta Z_{ZA}   + 2C_Z 
\eeqan
where, in the last line ${\cal Q} = 0.529$. 

Expansions of Higgs coupling parameters:
\beqa
\eta_W &=&  -\half c_H + 2 \delta m_W - \delta v \CR
\eta_Z  &=&  -\half c_H + 2 \delta m_Z - \delta v - c_T \CR
\eta_{ZZ}  &=&  -c_H + 2 \delta m_Z - 2\delta v - 5c_T \CR
\eta_h &=&  -\thalf c_H + c_6 + \delta\bar\lambda + \delta v\CR
\theta_h &=& c_H\CR
\zeta_W &=&  \delta Z_W \CR
\zeta_Z &=&  \delta Z_Z \CR
\zeta_A &=&  \delta Z_A \CR
\zeta_{AZ} &=&  \delta Z_{AZ} \CR
\eeqan

Expansions of effective $W$ vertex parameters:
\beqa
\delta g_{Z,eff} &=& \delta g_Z + {1\over c_w^2} ( (c_w^2 - s_w^2)
\delta g_L + s_w^2 \delta g_R - 2 \delta g_W) \CR
\delta \kappa_{A,eff} &=&  (c_w^2-s_w^2) (\delta g_L - \delta g_R) + 2
(\delta e -\delta g_W) + (8c_{WB}) \CR
\delta\lambda_{A,eff} &=& - 6 g^2 c_{3W} 
\eeqan

Expansions of $\ee\to Zh$ parameters:
\beqa
a_L &=& \eta_Z + \delta g_L + {(s - m_Z^2)\over 2m_Z^2} { (c_{HL} +
  c^\prime_{HL})\over (1/2 - s_w^2)} + k_Z\delta m_Z + k_h \delta m_h \CR     
a_R &=& \eta_Z + \delta g_R -  {(s - m_Z^2)\over 2m_Z^2} { c_{HE}
  \over s_w^2}+k_Z\delta m_Z + k_h \delta m_h \CR      \CR
b_L &=&  {1\over (1-2s_w^2)} \biggl\{  c_w^2 (1 - 2s_w^2 {m_Z^2\over
  s}) (8c_{WW}) +  2s_w^2 (1- 2s_w^2){m_Z^2\over s} (8c_{WB}) \CR & & 
   \hskip 0.2in -{s_w^4\over c_w^2} (1 - 2 c_w^2 {m_Z^2\over s}) (8c_{BB}) \biggr\}\CR
b_R &=& c_w^2 {m_Z^2\over s} (8c_{WW}) +  (1 - (1-2s_w^2){m_Z^2\over
  s}) (8c_{WB}) + {s_w^2\over c_w^2} (1 - c_w^2 {m_Z^2\over s}) (8
c_{BB}) 
\eeqan
In the formulae for $a_L$ and $a_R$, 
\beqa
k_Z &=& {2m_Z^2\over s-m_Z^2} + {E_Z m_Z^2\over 2k^2 \sqrt{s}} -
{m_Z^2\over 2 k^2} - {E_Z^2/m_Z^2\over (2 + E_Z^2/m_Z^2)} (1 -
{m_Z^2\over E_Z\sqrt{s}})\CR
k_h &=& - {E_Z m_h^2\over 2k^2 \sqrt{s}} -
{E_Z^2/m_Z^2\over (2 + E_Z^2/m_Z^2)} 
{m_h^2\over E_Z\sqrt{s}}
\eeqan

Expansions of $\sigma(\ee\to \nu\bar\nu h)$ for different CM energies:
\beqa
\delta\sigma(250) &=&  2\eta_W - 2\delta v + 2 \delta g_W -  1.6 \delta m_W - 3.7 \delta
m_h  \CR & & \hskip 0.2in  - 0.22 \,\delta Z_W - 6.4 c^\prime_{HL}- 0.37
  (c_{HL} - c^\prime_{HL}) \CR
\delta\sigma(350) &=&  2\eta_W - 2\delta v + 2 \delta g_W -  1.2 \delta
m_W - 2.0 \delta
m_h  \CR & & \hskip 0.2in  - 0.32  \,\delta Z_W - 7.5 c^\prime_{HL}- 0.28
  (c_{HL} - c^\prime_{HL}) \CR
\delta\sigma(380) &=&  2\eta_W - 2\delta v + 2 \delta g_W -  1.1 \delta
m_W - 1.7 \delta
m_h  \CR & & \hskip 0.2in  - 0.34  \,\delta Z_W - 7.8 c^\prime_{HL}- 0.26
  (c_{HL} - c^\prime_{HL}) \CR
\delta\sigma(500) &=& 2\eta_W - 2\delta v + 2 \delta g_W -  0.85 \delta
m_W - 1.2 \delta
m_h  \CR & & \hskip 0.2in  - 0.39  \,\delta Z_W - 8.8 c^\prime_{HL}- 0.19
  (c_{HL} - c^\prime_{HL}) 
\eeqan

Expansions of Higgs boson partial widths:
\beqa
\delta \Gamma(h\to b \bar b) & = &   - c_H + 2 c_{b\Phi} \CR
\delta \Gamma(h\to c \bar c) & = &   - c_H + 2 c_{c\Phi}\CR
\delta \Gamma(h\to \tau^+ \tau^-) & = &   - c_H + 2 c_{\tau\Phi} \CR
\delta \Gamma(h\to \mu^+\mu^-) & = &   - c_H + 2 c_{\mu\Phi} \CR
\delta \Gamma(h\to gg) & = &  - c_H + 2 c_{g\Phi}\CR
\delta \Gamma(h\to WW^*) &=& 2\eta_W - 2\delta v -  11.7 \delta
m_W + 13.6\delta
m_h  - 0.75 \delta Z_W - 0.88 C_W + 1.06
\delta \Gamma_W \CR
\delta \Gamma(h\to ZZ^*) &=& 2\eta_Z - 2\delta v -  13.8\delta
m_Z + 15.6\delta
m_h  - 0.50 \delta Z_Z - 1.02 C_Z + 1.18 
\delta \Gamma_Z \CR
\delta \Gamma(h\to \gamma\gamma) & = &  528  \,\delta Z_A  - c_H + 4
\delta e +  4.2 \, \delta m_h - 1.3  \,\delta m_W - 2 \delta v\CR
\delta \Gamma(h\to Z\gamma) &=& 290  \,\delta Z_{AZ} - c_H - 2(1-3
s_W^2)\delta g + 6 c_w^2 \delta g'+ \delta Z_A + \delta Z_Z\CR & &
\hskip 0.2in
+ 9.6 \, \delta m_h - 6.5 \, \delta m_Z -2 \delta v \CR
\eeqan

Expansions of $\sigma(\ee\to Zhh)$ at $\sqrt{s} = 500$~GeV for states
of given $\ee$ beam polarization: 
\beqa
\delta\sigma(L)  &=& 2 \delta g_L +  1.40 \eta_Z + 1.02 \eta_{ZZ} +
18.6 \delta Z_Z + 24.8 \delta Z_{AZ} \CR
  && \hskip 0.2in + 0.56 \eta_h - 1.58 c_H +  108.3 (c_{HL} +
  c^\prime_{HL}) \CR
  & & \hskip 0.2in  - 3.9 \delta m_h + 3.5 \delta m_Z  \CR
\delta\sigma(R) &=& 2 \delta g_R +  1.40 \eta_Z + 1.02 \eta_{ZZ} +
18.6 \delta Z_Z - 28.7 \delta Z_{AZ} \CR
  && \hskip 0.2in + 0.56 \eta_h - 1.58 c_H - 125.5 c_{HE}\CR
  & & \hskip 0.2in  - 3.9 \delta m_h + 3.5 \delta m_Z  \CR
\delta\sigma(U) &=& 1.15\delta g_L + 0.85 \delta g_R +   1.40 \eta_Z + 1.02 \eta_{ZZ} +
18.6 \delta Z_Z + 2.0 \delta Z_{AZ} \CR
  && \hskip 0.2in + 0.56 \eta_h - 1.58 c_H +  62.1  (c_{HL} +
  c^\prime_{HL}) - 53.5 c_{HE}\CR
  & & \hskip 0.2in  - 3.9 \delta m_h + 3.5 \delta m_Z  
\eeqan
In these equations $L$ refers to the beam polarization state $e^-_Le^+_R$,
$R$ refers to the beam polarization state $e^-_Le^+_R$, and $U$ refers to
unpolarized beams.  To  find the expressions for arbitrary
polarizations, it is useful to have the total cross sections for the
two completely polarized beam configurations:  $\sigma(L) = 0.36$~fb,
$\sigma(R) = 0.27$~fb.

\section{Values for projected uncertainties input into our analysis}

The 13-parameter fit described in Section 5 used as inputs projected uncertainties in
precision electroweak observables, LHC measurements of ratios of Higgs boson
branching ratios, and measurements of the $a$ and $b$ parameters of
$\ee\to Zh$ at the 500~GeV ILC.   For the precision electroweak
inputs, we have taken the values listed in Table~\ref{tab:PEW},
including the future improvements quoted there.   For LHC
measurements, we have used as our inputs
\beqa
   \delta (  BR(h\to ZZ^*)/BR(h\to \gamma\gamma) )  & = &   2\% \CR
   \delta (  BR(h\to Z\gamma)/BR(h\to \gamma\gamma) )  & = &  31\% \CR
   \delta (  BR(h\to \mu^+\mu^-)/BR(h\to \gamma\gamma) )  & = &   12\%
   \ .
\eeqan
as described in Section 3.   For the $a$ and $b$ parameter
measurements, we have used the estimates~\cite{Ogawa}
\beq
\begin{tabular}{lccc} 
beam polarization &   $\delta a$ & $ \delta b$ &  $ \rho(a,b)$ \\ \hline
-80\%/+30\%         &        4.0      &    0.70    &   84.8 \\
+80\%/-30\%         &        4.2      &    0.75    &   86.5 \\
\end{tabular}
\eeqn
with all numbers in \%. 

The final fit described in Section 8, which uses 22 parameters, makes use of a much larger number
of inputs.   These are listed in the Appedix of \cite{ImpHiggs}.  The
full set of linear relations given in Appendix A, and the final $22\times
22$ covariance matrices for the fit parameters given by  the ILC 250
fit and the full ILC fit are given in files {\tt
CandV250.txt}  and {\tt CandV500.txt} included with the arXiv posting of that  paper.

\section{Relation between the EFT and $S$, $T$ formalisms}

In the $S$, $T$ formalism for the interpretation of precision
electroweak measurements~\cite{PT}, we define a reference value of the
weak mixing angle from the quantities $\alpha(m_Z^2)$, $m_Z$, and $G_F$
and then compare the predictions for other precision electroweak
observables to expectations based on this value.   More specifically,
we define $\sin^2\theta_0$ by
\beq
    4 s_0^2 c_0^2 =    {4\pi \alpha\over \sqrt{2} G_F m_Z^2}  \ .
\eeqn
Then we can write expressions for precision electroweak observables in
terms of $s_0^2$.    The variations of the SM parameters conveniently cancel out of
these formulae in leading order.  For example,
\beqa
m_W^2/m_Z^2 &=&  c_0^2 +   {c_0^2 \over c_0^2 - s_0^2}  \bigl( c_0^2
\, c_T - 2 s_0^2 (c^\prime_{HL} + (8c_{WB}) \bigr)\CR
s_*^2 &=& s_0^2 + {s_0^2 \over c_0^2 - s_0^2} \bigl(c^\prime_{HL} + (8c_{WB})
- c_0^2 c_T \bigr) - \half c_{HE} - s_0^2 (c_{HL} - c_{HE} ) \ ,
\eeqa{PEWST}
where $s_*^2$ is the value of the weak mixing angle that governs the
polarization asymmetries at the $Z$ pole.

The $S$ and $T$ parameters are defined so that, in the approximation
in which all precision electroweak corrections arise from vacuum
polarization diagrams, the formulae \leqn{PEWST} take the form
\beqa
m_W^2/m_Z^2 &=&  c_0^2 +   {\alpha c_0^2 \over c_0^2 - s_0^2}  \bigl(
-\half S + c_0^2\, T \bigr)
\CR
s_*^2 &=& s_0^2 + {\alpha \over c_0^2 - s_0^2} \bigl( {1\over 4} S -
s_0^2 c_0^2 T ) 
\eeqa{PEWSTST}
Then we can identify
\beqa
   \alpha\  S & = & 4 s_0^2 ( 8c_{WB} + c^\prime_{HL} )\CR
  \alpha\ T &= & c_T \ .
\eeqan

The $S$, $T$ formalism was quite appropriate for the experimental
situation of the early 1990's, when $\alpha$, $G_F$, and $m_Z$ were by
far the best-measured electroweak parameters.   Today, the
uncertainties in  $m_W$ and
$A_\ell$ have improved to the point where these observables 
 should be treated on the same footing. The formalism used in this
 paper is more democratic with respect to possible choices of
 the reference electroweak parameters.


\begin{thebibliography}{99}
\bibitem{Disclaimer}
We ignore the logically possible but highly fine-tuned hypothesis that
the new physics contribution to these decays is almost exactly $(-2)$
times the SM contribution.
\end{thebibliography}
\end{document}